\documentclass[preprint, trackchanges]{aastex63}
\usepackage{multirow}
\usepackage{amsmath}
\usepackage{appendix}

\shorttitle{Data-driven simulation of the flux rope}

\begin{document}

\title{Data-driven Modeling of a Coronal Magnetic Flux Rope: from Birth to Death}

\correspondingauthor{P. F. Chen and Y. Guo}
\email{chenpf@nju.edu.cn, guoyang@nju.edu.cn}

\author[0000-0002-4205-5566]{J. H. Guo}
\affiliation{School of Astronomy and Space Science and Key Laboratory for Modern Astronomy and Astrophysics, Nanjing University, Nanjing 210023, China}

\affiliation{Centre for mathematical Plasma Astrophysics, Department of Mathematics, KU Leuven, Celestijnenlaan 200B, B-3001 Leuven, Belgium}

\author[0000-0002-9908-291X]{Y. W. Ni}
\affiliation{School of Astronomy and Space Science and Key Laboratory for Modern Astronomy and Astrophysics, Nanjing University, Nanjing 210023, China}

\author[0000-0002-9293-8439]{Y. Guo}
\affiliation{School of Astronomy and Space Science and Key Laboratory for Modern Astronomy and Astrophysics, Nanjing University, Nanjing 210023, China}

\author[0000-0002-7153-4304]{C. Xia}
\affiliation{School of Physics and Astronomy, Yunnan University, Kunming 650500, China}

\author[0000-0003-3364-9183]{B. Schmieder}
\affiliation{Centre for mathematical Plasma Astrophysics, Department of Mathematics, KU Leuven, Celestijnenlaan 200B, B-3001 Leuven, Belgium}
\affiliation{LESIA, Observatoire de Paris, CNRS, UPMC, Universit\'{e} Paris Diderot, 5 place Jules Janssen, F-92190 Meudon, France}

\author[0000-0002-1743-0651]{S. Poedts}
\affiliation{Centre for mathematical Plasma Astrophysics, Department of Mathematics, KU Leuven, Celestijnenlaan 200B, B-3001 Leuven, Belgium}
\affiliation{Institute of Physics, University of Maria Curie-Skłodowska, ul.\ Radziszewskiego 10, 20-031 Lublin, Poland}

\author[0000-0001-5483-6047]{Z. Zhong}
\affiliation{Center for Integrated Research on Space Science, Astronomy, and Physics, Institute of Frontier and Interdisciplinary Science, Shandong University, Qingdao 266237, China}

\author[0000-0002-4391-393X]{Y. H. Zhou}
\affiliation{Centre for mathematical Plasma Astrophysics, Department of Mathematics, KU Leuven, Celestijnenlaan 200B, B-3001 Leuven, Belgium}

\author[0000-0002-1713-2160]{F. Yu}
\affiliation{Key Laboratory of Dark Matter and Space Astronomy, Purple Mountain Observatory, Chinese Academy of Sciences, Nanjing 210023, People’s Republic of China}
\affiliation{School of Astronomy and Space Science, University of Science and Technology of China, Hefei 230026, People’s Republic of China}

\author[0000-0002-7289-642X]{P. F. Chen}
\affiliation{School of Astronomy and Space Science and Key Laboratory for Modern Astronomy and Astrophysics, Nanjing University, Nanjing 210023, China}

\begin{abstract}
Magnetic flux ropes are a bundle of twisted magnetic field lines produced by internal electric currents, which are responsible for solar eruptions and are the major drivers of geomagnetic storms. As such, it is crucial to develop a numerical model that can capture the entire evolution of a flux rope, from its birth to death, in order to predict whether adverse space weather events might occur or not. In this paper, we develop a data-driven modeling that combines a time-dependent magneto-frictional approach with a thermodynamic magnetohydrodynamic model. Our numerical modeling successfully reproduces the formation and confined eruption of an observed flux rope, and unveils the physical details behind the observations. Regarding the long-term evolution of the active region, our simulation results indicate that the flux cancellation due to collisional shearing plays a critical role in the formation of the flux rope, corresponding to a substantial increase in magnetic free energy and helicity. Regarding the eruption stage, the deformation of the flux rope during its eruption can cause an increase in the downward tension force, which suppresses it from further rising. This finding may shed light on why some torus-unstable flux ropes lead to failed eruptions after large-angle rotations. Moreover, we find that twisted fluxes can accumulate during the confined eruptions, which would breed the subsequent eruptive flares.

\end{abstract}


\section{Introduction} \label{sec:intro}

A magnetic flux rope (MFR) is a fundamental magnetic configuration in the solar atmosphere and interplanetary space, defined as a bundle of field lines winding around a common axis. In remote-sensing imaging observations of the solar atmosphere, flux ropes can be traced and characterized by filaments/prominences, coronal cavities, sigmoids, and hot channels, and are generally regarded as the key for coronal mass ejections (CMEs) and flares \citep{Chen2011, Schmieder2013, Chen2017}. Once a magnetic flux rope erupts and escapes into interplanetary space, it is termed an interplanetary coronal mass ejection \citep[ICME;][]{Burlaga1981}, and its impact on the magnetosphere jeopardizes navigation and communication satellites. Accordingly, the studies for the birth, ejection, and propagation of flux ropes are extremely important to understand solar eruptions and predict hazardous space weather events. Albeit a flux rope has been recognized as the core of a solar eruption for many years, the underlying mechanisms concerning its formation and eruption are still under debate \citep{Patsourakos2020}. For instance, how is it formed: before or during the eruption? Why do some flux ropes remain confined in the solar atmosphere, failing to evolve into a CME?

Regarding their birth, while it was frequently claimed that flux ropes are crucial for eruptions, some researchers have found that a flux rope is not necessary prior to eruption, and it can be formed during the eruption. For example, \citet{Song2014} and \citet{Wangws2017} traced the evolution of flux rope structures and found that they are formed via reconnection during eruption but not before. With magnetohydrodynamic (MHD) simulations, \citet{Jiang2021} confirmed the ``tether-cutting" model \citep{Moore1980}, i.e., an eruption results from reconnection between two J-shaped arcades, and the flux rope is formed during the eruption. Statistical studies showed that approximately 89\% of eruptive filaments are supported by flux ropes before eruption, and the remaining 11\% are sheared arcades, indicating that flux ropes are dominant, but not necessary, in the pre-eruptive structures \citep{Ouyang2017}. Note that there are cases where a sheared arcade and a flux rope co-exist in a filament \citep{Guo2010}. The reason for the high percentage is that flux ropes can set the stage for instabilities to trigger eruptions, such as the kink instability \citep{Torok2004} and the torus instability \citep {Kliem2006}, where the former is evaluated via the twist number of the flux rope and the latter via the decay index of the external magnetic field above the flux rope. Additionally, flux ropes are effective structures for storing nonpotential energy. Moreover, it is much easier for a current sheet to be formed below a flux rope, whose reconnection leads to a final eruption \citep{Chen2011}. Regarding the eruptiveness, after being triggered to rise by either ideal MHD instabilities or induced by magnetic reconnection \citep{Welsch2017}, the flux rope may escape into interplanetary space or remain confined in the solar atmosphere and turn into a failed eruption \citep{Ji2003}. Although these scenarios are generally accepted and investigated by many numerical simulations under ideal configurations \citep{Fan2009, Aulanier2010}, the complexities of mechanisms and processes in real situations make them still worthy of exploring with simulations employing/ingesting observational data.

Recently, numerical models using observational data as inputs have shown tremendous potential in unraveling the complexity of the magnetic topology and thermodynamic evolution behind observations \citep{Jiang2022}. These models use observed photospheric magnetograms (or derived products) to reproduce the evolution of magnetic fields and thermodynamics in the corona. They generally fall into two categories: data-constrained and data-driven types. In the former, the initial magnetic field is usually reconstructed using the nonlinear force-free field (NLFFF) extrapolation method, and the photospheric or low-coronal boundary is subsequently either fixed or provided by numerical extrapolations. This means that subsequent computations are no longer influenced by the continuous input of observational data. Data-constrained simulations are mainly utilized to reproduce eruption events \citep{Kliem2013, Inoue2018nc, Jiang2018, Guoy2021, Guojh2023b}, in which the coronal magnetic field undergoes significant changes in minutes, making the effects of boundary evolution negligible. On the other hand, data-driven models adopt a time-series of magnetic field, velocity field and/or electric field in the photosphere as the input data to keep the computation synchronized with the observations at every time step. Therefore, the evolution of the coronal magnetic fields and plasma is directly driven by the observational data. Compared to data-constrained types, data-driven models can capture the real-time responses of the corona to the photosphere and are suitable for studying the long-term evolution of active regions \citep{Cheung2012, Jiang2016, Hayashi2018, Pomoell2019, Inoue2023, Chen2023}. 

In our previous work, we utilized the data-driven technique to reproduce some observational features of eruption events, where the initial magnetic field is given by the NLFFF model and the driven duration is within two hours \citep{Guo2019, Zhong2021, Guo2023}. However, the flux rope formation process and the active-region long-term evolution were not incorporated. To exhibit the advantages of the data-driven technique and study the entire process of the flux rope from its birth to eruption, we couple the time-dependent magneto-frictional (TMF) approach with the data-driven thermodynamic MHD model \citep{Guo2023} in this paper. The initial magnetic field, well before eruption, is a potential field, and the subsequent coronal evolution is fully driven by the observational data in the photosphere via the TMF approach. The final state of this modeling further serves as the initial condition for the thermodynamic MHD simulation, allowing us to explore the formation, eruption, and confined mechanisms of the observed flux rope. The event overview and numerical strategies are described in Sections~\ref{sec:obs} and \ref{sec:method}, respectively. The numerical results are presented in Section~\ref{sec:results}, which are followed by a discussion in Section~\ref{sec:discussion}. Finally, we summarize our findings in Section~\ref{sec:summary}.

\section{Observation overview} \label{sec:obs}

The active region we study in this paper is NOAA Active Region 12673, which was the most flare-productive active region in solar cycle 24, hosting four X-class and 27 M-class flares from 2017 September 4 to 10 \citep{Yang2017}. Its on-disk evolution was well-observed by the Atmospheric Imaging Assembly \citep[AIA;][]{Lemen2012} and the Helioseismic and Magnetic Imager \citep[HMI;][]{Scherrer2012} on board the Solar Dynamics Observatory (SDO). On 2017 September 6, a confined X2.2 flare occurred at 08:57 UT when the active region was relatively well developed, followed by a successive X9.3 flare at 11:53 UT accompanied by a CME approximately three hours later \citep{Liu2018}. 

To understand why this active region is so productive, how magnetic energy accumulates in it, and what leads to the difference between these homologous eruptions, many observations and MHD simulations using observational data as inputs have been conducted. For instance, \citet{Sun2017} demonstrated that magnetic fluxes in this active region emerge faster than those in any previously observed active region. \citet{Yan2018} observed that this active region exhibits rotational motions near the flaring site. \citet{Liu2019} found that shearing flows and flux cancellation between opposite polarities are responsible for the formation of flux ropes. Many static NLFFF extrapolations and MHD simulations showed that the two successive X-class flares stem from the eruptions of flux ropes \citep{Inoue2018, Liu2018, Jiang2018, Hou2018, Liu2019, Price2019, Inoue2021}. Studies conducted by \citet{Moraitis2019} and \citet{Price2019} investigated the evolution of magnetic helicity using continuous NLFFF extrapolations and data-driven simulations, respectively. Both studies suggested that the helicity ratio is effective in predicting eruptivity. \citet{Wangr2018} found that the majority of helicity is built up by the shearing and converging flows acting upon pre-existing and emerging fluxes. Additionally, \citet{Scolini2020} revealed that the CMEs produced by this active region can interact with each other and form complex ejection in the heliosphere, leading to an intense geomagnetic storm. Hence, this active region provides a valuable opportunity to study the entire timeline of flux rope development from birth to eruption.

In this paper, we aim to address several questions related to NOAA active region 12673, using our newly developed data-driven model. Specifically, we seek to reproduce the long-term evolution of this active region, the timeline of the flux rope development, and the confined nature of the X2.2 eruption (SOL2017-09-06T08:57), with the purpose to answer the following questions: (1)~How is the flux rope formed? (2)~Why is the X2.2 flare confined? (3)~Is the first confined X2.2 flare related to the following eruptive X9.3 flare three hours later?

\section{Modeling description} \label{sec:method}

In general, the occurrence of a solar flare involves two primary phases: the long-term accumulation of nonpotential energy during the buildup stage, followed by a drastic release of magnetic energy in the eruption process. It is well accepted that they exhibit significant differences in timescales. As estimated in \citet{Demoulin2010}, in a well-developed active region with a flux of $10^{22}\;$Mx, the substantial alteration of the coronal currents and the normal magnetic fields in the photosphere requires a prolonged period ranging from a few days to several weeks. However, the subsequent coronal eruption process is more drastic, in which the magnetic configuration significantly changes in just tens of seconds or minutes due to magnetic reconnection. It should be noted that magnetic fluxes emerge more rapidly in NOAA active region 12673 than normally in active regions \citep{Sun2017}, potentially leading to a departure from quasi-static evolution. Nevertheless, the release of magnetic energy in the eruption process is typically faster than its buildup. Hence, to effectively capture these distinct processes with varying timescales while optimizing computational resources, we opt for the combination of the following two models: the TMF model and the thermodynamic MHD model. The former is used to describe the long-term buildup phase (\S\ref{sec:for}), and the latter is employed to simulate the drastic eruption process (\S\ref{sec:eru}). This hybrid model has been demonstrated to be effective in reproducing observed solar eruptions \citep{Afanasyev2023, Wagner2023, Daei2023}. The simulation is fully driven by the photospheric observational data, which are detailed in Section~\ref{sec:boundary}.

The partial differential equations in both aforementioned models are numerically solved with the Message Passing Interface Adaptive Mesh Refinement Versatile Advection Code \citep[MPI-AMRVAC\footnote{http://amrvac.org},][]{Xia2018, Keppens2023}. The computational domain is $[x_{min},x_{max}]\ \times [y_{min},y_{max}]\ \times [z_{min},z_{max}] = [-102.6, 102.6]\ \times [-73.3,73.3]\ \times[1,205.2]\;\rm Mm^{3}$, with the effective mesh grid of $560 \times 400 \times 560$, adopting a four-level adaptive mesh refinement. Due to the refinement technique, the grid spacing can attain the limiting pixel size of SDO/HMI data, $0\farcs 5$. 

The AMRVAC computational framework employs the finite-volume method for the solution of the MHD equations. The fundamental procedure involves the reconstruction of variables at cell faces through a third-order limiter \citep{Cada2009}. Subsequently, the calculation of corresponding fluxes is performed utilizing the classic HLL scheme \citep{Harten1983}, conforming to two ghost cells on each boundary. To mitigate the divergence of magnetic fields arising during the numerical computation, we use the divergence-cleaning method described in \citet{Keppens2003} in the TMF model, and the generalized Lagrange multiplier \citep[GLM;][]{Dedner2002, Mignone2010} in the MHD model. To evaluate the solenoidality of the simulated magnetic fields, we calculate the divergence-free metric proposed by \citet{Wheatland2000}, namely, $\small{<}|f_{i}|\small{>}$, which describes the volume-weighted average of the absolute value of the fractional magnetic flux. Over the approximate 80-hour simulation duration, the flux variation remains consistently within the range of $2.0 \times 10^{-4}$, with the minimum and maximum values being $5.77 \times 10^{-5}$ and $1.62 \times 10^{-4}$, respectively. These values are generally considered acceptable in previous NLFFF extrapolations \citep{Valori2013, Guoy2016a, Thalmann2019}. Moreover, as demonstrated by \citet{Thalmann2019}, the non-solenoidality energy ratio is less than 0.05 in cases where $\small{<}|f_{i}|\small{>}$ is smaller than $5 \times 10^{-4}$. This indicates that our computation of magnetic fields is credible.

\subsection{Data-driven boundary conditions} \label{sec:boundary}

We utilize time-dependent observational data to drive the coronal magnetic field evolution. Following \citet{Guo2019}, the vector magnetic field and the derived velocity fields in the photosphere are implemented together at the bottom boundary, which is called $v$--$B$ driven model. The Space-weather HMI Active Region Patches (SHARP) vector magnetograms \citep[hmi.sharp\_cea\_720 s;][]{Bobra2014} are utilized in our paper. They are remapped onto a spherical coordinate system with the cylindrical equal-area (CEA) projection centered at the tracked active region. Besides, the vector magnetic field of the SHARP data have undergone preprocessing steps \citep{Bobra2014}, such as resolving the $180^{\circ}$ ambiguity of the azimuthal component \citep{Metcalf1994, Leka2009}, noise reduction \citep{Couvidat2012}, and active region tracking. These advantages determine that the SHARP data are well-suitable for data-driven simulations, as demonstrated by \citet{Jiang2016}. The photospheric flows are derived from the Differential Affine Velocity Estimator for Vector Magnetograms \citep[DAVE4VM;][]{Schuck2008} method.

Despite the aforementioned preprocessing steps that have been applied to the SHARP data, the original magnetograms still retain a notable level of noise, which could lead to small-scale fluctuations during the numerical calculations. To mitigate this issue, similar to our previous works \citep{Guo2019, Zhong2021, Guo2023}, we employ additional preprocessing proposed by \citet{Wiegelmann2006} to the temporal sequence of the vector magnetograms. This approach utilizes an optimization method to minimize the function composed of four terms (Equation (6) in \citeauthor{Wiegelmann2006}~\citeyear{Wiegelmann2006}), including the deviation from the observational data, smoothness of the magnetic fields, Lorentz force and torque. By executing this approach, the Lorentz force and torque can be effectively reduced to a small value, achieving smoothness of the vector magnetic field within the measurement error. In practice, we perform 5000 iterations for each magnetogram, after which the Lorentz force and torque decrease to one thousandth of the original magnitudes (quantified by $\epsilon_{force}$ and $\epsilon_{torque}$ in \citeauthor{Wiegelmann2006}~\citeyear{Wiegelmann2006}), and the magnetic fields are smoothed. It is worth noting that, unlike NLFFF extrapolations, there is no strict requirement to satisfy the force-free condition for data-driven simulations of the TMF model. Nevertheless, we also employ this technique to preprocess the input vector magnetograms due to its positive impact on enhancing the numerical stability according to our previous numerical tests. Hereafter, the time series of processed vector magnetograms is fed into the DAVE4VM code to derive the velocity fields in the photosphere. We choose a window size of 19 pixels in the DAVE4VM, consistent with previous works \citep{Lumme2017, Jiang2021F}. In addition, as the provided observational data are available only at discrete time intervals with a cadence of 12 minutes, we adopt a linear interpolation in time to fill in the missing data between these intervals.

We employ the $v$--$B$ driven boundary to evolve the coronal magnetic fields, which is different from the electric field data-driven ($E$-driven hereafter) models \citep{Cheung2012, Pomoell2019}. In $E$-driven models, the first purpose is to reproduce the magnetic-field evolution in observations with the derived electric fields. On the other hand, the $v$--$B$ driven models directly inputting both the magnetic fields and velocities at the bottom boundary. Although such boundary conditions seem over-prescribed to comply with the MHD equations, its effectiveness in reproducing solar eruptions has been demonstrated by many works \citep{Guo2019, Liuca2019, He2020, Zhong2021, Guo2023, Zhong2023}. Note that the velocity field derived via the DAVE4VM method, cannot precisely reproduce the evolution of the vector magnetic field in observations due to the fact that the DAVE4VM electric field ($\boldsymbol{-v_{_{\rm DAVE4VM}} \times B}$) fulfills Faraday's law only in the least squares sense \citep{Schuck2008}, thereby leading to a partial loss of inductive property when retrieving observed magnetograms \citep{Lumme2017}. The adoption of the DAVE4VM velocity field for specifying boundary conditions in data-driven simulations remains questionable. Given this, the incorporation of magnetic fields may be an effective complement to the $v$-driven boundary, where only the velocity field is used as the boundary condition. This can potentially enhance the precision of magnetic-field evolution, constraining it closer to observations.

The vector magnetic field and velocity field are directly imposed at the cell centers of the inner ghost layer (closer to the physical domain). Consequently, the photosphere ($z=0$) serves as the boundary condition rather than being incorporated within the computational/physical domain. The values on the outer ghost cells are obtained through a zero-gradient extrapolation. This approach ensures that the continuous change of the bottom boundary condition of the simulation remains consistently synchronized with observations. Following this, the evolution within the physical domain can be computed using the aforementioned numerical scheme, driven by the input of observational data.

\subsection{TMF stage: long-term quasi-static evolution of the active region} \label{sec:for}

Compared to the rapid eruption that occurs within minutes, the pre-eruptive buildup process, spanning several days, manifests a significantly slower evolution. Moreover, owing to the difference in the Alfv\'en speed and dynamic timescales between the corona and the photosphere, the coronal magnetic field can rapidly respond to changes in the photosphere. Therefore, from the viewpoint of the photosphere, the corona can be considered as a series of quasi-static states. To obtain the continuous evolution of the three-dimensional (3D) coronal magnetic field in a fast way, the TMF method has been proposed \citep{Cheung2012, Pomoell2019}, which evolves the magnetic field to a force-free state by solving the magnetic induction equation, where the velocity is proportional to the local Lorentz force. It is worth noting that the equation of the TMF model is parabolic \citep{Craig1986}, such that a stable step in the explicit time-stepping scheme is required. In practice, the maximal signal speed ($C_{_{\rm max}}$) is determined as $v_{_{\rm MF}}+v_{_{\rm A}}$, where $v_{_{\rm MF}}$ represents the magneto-frictional velocity and $v_{_{\rm A}}$ is the local Alfv\'en speed. Notably, the Alfv\'en speed across the entire simulation domain is set to a uniform value of 1 in the normalization unit. Thereafter, the adaptive time step is given by $\Delta t={\rm min}\{C_{\scriptscriptstyle \rm CFL} \frac{\Delta x}{C_{_{\rm max}}}, \frac{\Delta x^{2}}{2\eta_{_{\rm max}}}\}$, where the constant $C_{\scriptscriptstyle \rm CFL}$ is 0.8 in our simulation.

The evolution of the coronal magnetic field is driven by the time series of observational data in the photosphere. This approach allows for a faster computation compared to the full MHD model while retaining the capability to capture the evolution of 3D magnetic fields. Similar to previous works \citep{Cheung2012, Pomoell2019, Price2019}, we investigate the long-term evolution of active region 12673 using the TMF model. The governing equations are as follows:

\begin{eqnarray}
 && \frac{\partial \boldsymbol{B}}{\partial t} + \nabla \cdot(\boldsymbol{vB-Bv})= -\nabla \times(\eta \boldsymbol{j}),\label{eq1}\\
 && \boldsymbol{v}=\frac{1}{\nu}\ \frac{\boldsymbol{j \times B}}{ B^{2}},\label{eq2}\\
  && \nu= \frac{\nu_{0}}{1-e^{-z/L}},\label{eq3}\\
 \notag  
 \end{eqnarray}
where $\nu_0=10^{-15}\;$s\;cm$^{-2}$ is the viscous coefficient of the friction, $L=10\;$Mm is the decay spatial scale of the viscosity toward the boundaries and $\eta$ is the magnetic diffusivity, which are employed by \citet{Cheung2012} and \citet{Pomoell2019}. In this work, we employ an anomalous resistivity $\eta$, which is defined as follows:

\begin{eqnarray}
  && \eta= \eta_{0}+\eta_{1}\frac{50\zeta}{1+e^{-2(\zeta-3.0)}},\label{eq4}\\
 \notag  
 \end{eqnarray}
where $\eta_{0}=2\times10^{11}\ \rm cm^{2}\ s^{-1}$, $\eta_{1}=2\times10^{12}\ \rm cm^{2}\ s^{-1}$, $\zeta=j^{2}B^{-2}\triangle^{2}$, and $\triangle=$min$[\triangle x, \triangle y, \triangle z]$. The anomalous resistivity is composed of two terms. The first term is a constant with a uniform distribution throughout the entire simulation domain, which is used to enhance the numerical stability. The second term is designed to enhance diffusion in the current sheet characterized by $J/B$. The initial magnetic field is a potential field, which is extrapolated using the Green's function method \citep{Chiu1977}. We set the bottom boundary of the potential field equal to the $B_z$ component of the photospheric vector magnetogram at 09:00 UT on 2017 September 3, and lateral and top boundaries are not involved. This suggests that the free energy is fully injected from the bottom-driven boundary due to the ensuing evolution. With the aid of this modeling, we simulate the long-term evolution of active region 12673 from 09:00 UT on 2017 September 3 to 13:00 UT on 2017 September 6, i.e., for about three days, involving more than 300 magnetograms. 

It should be noted that, the TMF model used in this paper, differs from the static magneto-frictional model employed for NLFFF extrapolation described in \citet{Guo2016b}. First, the TMF model encompasses the historical evolution of the coronal magnetic field over time, in contrast to the static magneto-frictional model that calculates the coronal magnetic field by inputting the photospheric magnetogram at a given time only. Second, the treatment of the boundary condition also varies between the two models. In the TMF model, the bottom boundary changes continuously and is driven by a time series of vector magnetograms and velocity fields, making it classified as a data-driven type. On the other hand, the static magneto-frictional model relies on a single magnetogram, hence referred to as the static NLFFF extrapolation. Both of these models were incorporated in MPI-AMRVAC 3.0 \citep{Keppens2023}. 

\subsection{MHD stage: rapid eruption} \label{sec:eru}

Different from the quasi-static evolution of the flux-rope formation that occurs on timescales of several days, the flux rope eruption involves numerous intricate physical processes, such as magnetic reconnection, heating, and waves. These processes can change the magnetic topology and thermodynamic properties in a few minutes. Therefore, to self-consistently reproduce the eruption features, a more realistic and complex model considering thermal source and sink terms is needed. We employ the data-driven thermodynamic MHD model developed in our previous work \citep{Guo2023}, whose input magnetic field is inherited from the TMF model at 08:56 UT on 2017 September 6. It is noticed that the magnetic fields in the TMF model evolves more slowly compared to those in thermodynamic MHD model. As a result, the choice of the switching moment may potentially influence the results of the MHD simulation. We select the magnetic fields just at the flare onset time in this paper. To see the impacts of the switching moment, the readers are referred to \citet{Daei2023}. The data-driven boundary is consistently applied throughout the entire MHD simulation, as in the TMF stage. The governing equations of the thermodynamic MHD model are as follows:

\begin{eqnarray}
 && \frac{\partial \rho}{\partial t} +\nabla \cdot(\rho \boldsymbol{v})=0,\label{eq5}\\
 && \frac{\partial (\rho \boldsymbol{v})}{\partial t}+\nabla \cdot(\rho \boldsymbol{vv}+p_{_{tot}}\boldsymbol{I}-\frac{\ \boldsymbol{BB}}{\mu_{0}})=\rho \boldsymbol{g},\label{eq6}\\
 && \frac{\partial \boldsymbol{B}}{\partial t} + \nabla \cdot(\boldsymbol{vB-Bv})=0,\label{eq7}\\
 && \frac{\partial \varepsilon}{\partial t}+\nabla \cdot(\varepsilon \boldsymbol{v}+p_{_{tot}}\boldsymbol{v}-\frac{\boldsymbol{BB}}{\mu_{0}}\cdot \boldsymbol{v}) =\rho \boldsymbol{g \cdot v}+H_{0}e^{-z/\lambda}-n_{\rm e}n_{\rm _{H}}\Lambda(T) \\
 \notag                                                       
 && + \nabla \cdot(\boldsymbol{\kappa} \cdot \nabla T), \label{eq8} 
\end{eqnarray}
where $p_{tot} \equiv p + B^2 / (2\mu_{0})$ is the total pressure with the postulation of full ionization, $\boldsymbol{g}=-g_{\odot}r_{\odot}^2/(r_{\odot}+z)^2\boldsymbol{e_{z}}$ is the gravitational acceleration, $g_{\odot}= \rm 274\ m\ s^{-2}$ is the gravitational acceleration at the solar surface, $r_{\odot}$ is the solar radius, $\varepsilon =\rho v^2/2+p/(\gamma -1)+ B^2 / (2\mu_{0})$ is the total energy density, $\boldsymbol{\kappa}=\kappa_{\parallel}\boldsymbol{\hat{b}\hat{b}}$ represents the field-aligned thermal conduction, $\kappa_{\parallel} =10^{-6}\ T^{\frac{5}{2}}\ \rm erg\ cm^{-1}\ s^{-1}\ K^{-1}$ is the Spitzer heat conductivity, $n_{\rm e}n_{\rm _{H}}\Lambda(T)$ is the optically-thin radiative losses, $H_{0}e^{-z/\lambda}$ is an empirical heating to maintain the high temperature of the corona, $H_0=10^{-4} \rm \ erg \ cm^{-3} \ s^{-1}$, $\lambda=60\ \rm Mm$, and the other parameters have their usual meanings. 

Regarding the initial atmosphere, we adopt a hydrostatic atmospheric model from the chromosphere to the corona, as follows:
\begin{eqnarray}
 T(z)=  \begin{cases} T_{_{\rm ch}}+\frac{1}{2}(T_{_{\rm co}}-T_{_{\rm ch}})({\rm tanh}(\frac{z-h_{_{\rm tr}}- 0.27}{w_{\rm tr}})+1)\qquad&  z \leq h_{_{tr}}, \\ (\frac{7}{2}\frac{F_{\rm c}}{\kappa}(z-h_{\rm tr})+T_{_{\rm tr}}^{7/2})^{2/7} & z > h_{_{tr}} \end{cases} \label{eq9}
 \end{eqnarray}
where $T_{_{\rm ch}}=8000\;$K denotes the chromospheric temperature, $T_{_{\rm co}}=1.5\;$MK is the coronal temperature, $h_{_{\rm tr}}=2\ $Mm and  $w_{_{\rm tr}}=0.2\;$Mm determine the height and thickness of initial transition region, and $F_{\rm c}=2 \ \times 10^{5}\ \rm erg \ cm^{-2}\ s^{-1}$ is the constant thermal conduction flux. Subsequently, the density distribution is computed from the bottom where the number density is set to be $1.15 \times 10^{15}\rm \ cm^{-3}$. To improve numerical efficiency and reduce numerical dissipation, similar to previous works \citep{Jiang2016, Kaneko2021, Guo2023}, the magnetic-field strength in this model is scaled down by a factor of 15. 
The manipulation of decreasing magnetic fields results in the values of plasma $\beta>1$ as the height exceeds 80~Mm. Nevertheless, the minimum value of plasma $\beta$ in the computation domain is about $2\times10^{-4}$, corresponding to the largest Alfv\'enic speed $v_{_{\rm A}}$ of approximately 8000~km s$^{-1}$. The conditions of plasma $\beta<1$ and $v_{_{\rm A}}>v_{_{\rm S}}$ are still satisfied within the regions relevant to the eruption. It means that the eruption process is still mainly dominated magnetically since the simulated flux rope is almost halted at 40 Mm for the confined eruption (see \S\ref{ce}).

\section{Numerical Results} \label{sec:results}

\subsection{Long-term evolution of the active region and formation of the flux rope} \label{sec:le}

\subsubsection{Global picture of the active region evolution} 
\begin{figure}[ht!]
\centering
\includegraphics[scale=0.13]{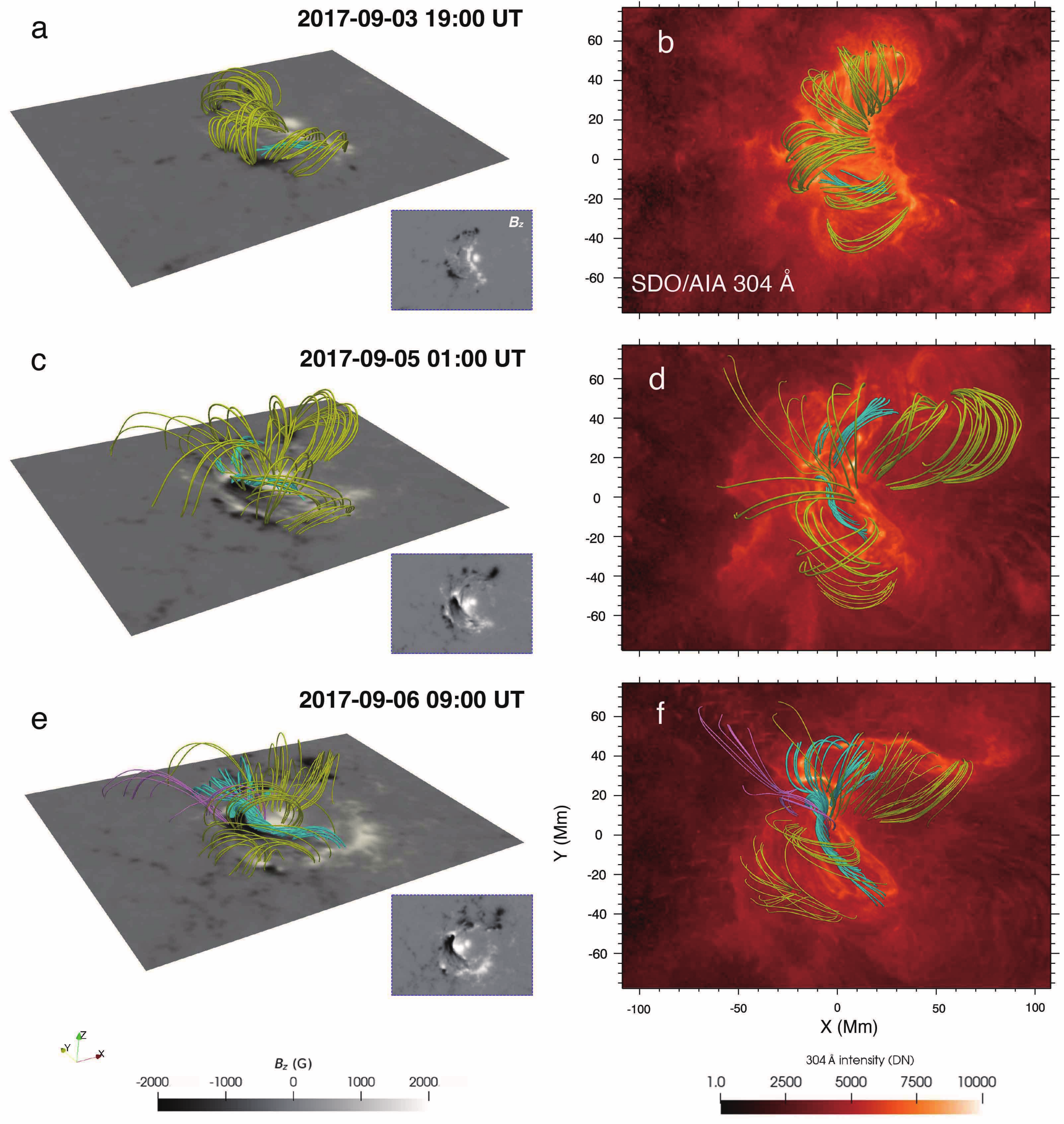}
\caption{Snapshots of the 3D evolution of the coronal magnetic field at 19:00 UT on 2017 September 3 (panels a and b), 01:00 UT on 2017 September 5 (panels c and d) and 09:00 UT on 2017 September 6 (panels e and f). The lines colored in cyan, yellow, and purple represent the highly sheared or twisted magnetic fields, background fields, and null-fan-spine structure, respectively. The insert figures display the distributions of the $B_{z}$ component on the bottom layer of the simulation domain. The left column shows a side view overlaid on the $B_z$ magnetograms, and the right column shows a top view overlaid on the SDO/AIA 304 \AA\ images.}
\label{figure1}
\end{figure}

Figures~\ref{figure1}a, \ref{figure1}c, and \ref{figure1}e show the 3D magnetic field evolution overlaid on the normal magnetic fields on the bottom layers, where the photospheric magnetic fields are from observations. From 19:00 UT on 2017 September 3 to 09:00 UT on 2017 September 6, the magnetic-field distributions on the bottom almost transform from a simple bipole to a multipolar configuration. During this period, new conjugate polarities emerge in succession, increasing the total unsigned flux by approximately 300\%. These emerging conjugate polarities almost simultaneously separate from each other and collide with the pre-existing polarities, forming a new polarity inversion line (PIL) referred to as the central PIL herein. More details about the evolution of the magnetic fields and flows in the photosphere can be found in \citet{Liu2019}.

\begin{figure}[ht!]
\centering
\includegraphics[scale=0.42]{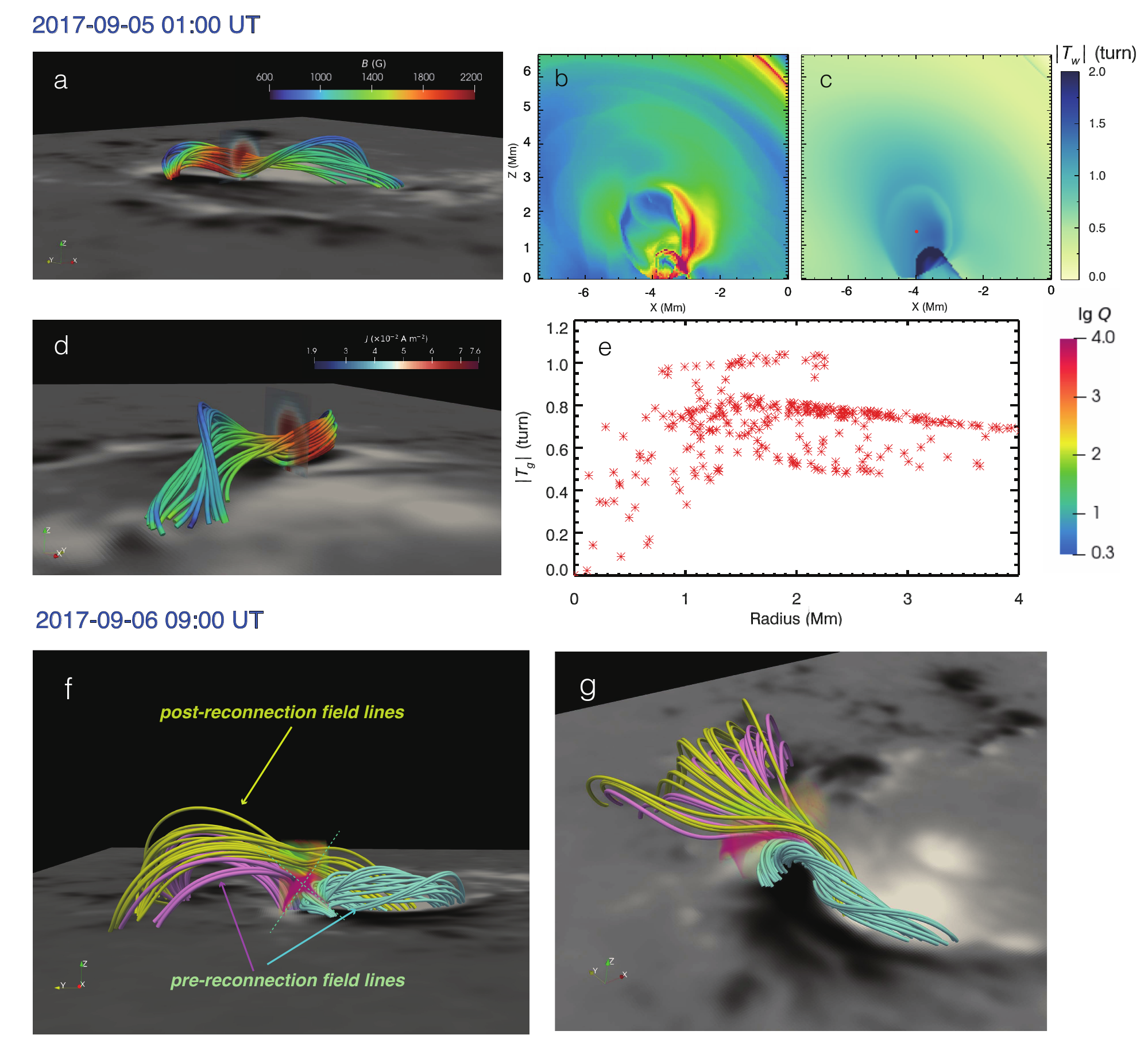}
\caption{Magnetic topological structure of the flux rope at 01:00~UT on 2017 September 5 (a, b, c, d, e) and 09:00~UT on 2017 September 6 (f, g). Panels (a) and (d) illustrate the side and end views of the typical field lines of the flux rope at 01:00~UT on 2017 September 5, respectively. The semi-transparent vertical slices across the flux rope showcase the distributions of the total electric current density. Panels (b) and (c) present the squashing factor $Q$ and $T_{w}$ maps on the same planes as Panels (a) and (d), where the red dot denotes the selected flux rope axis that is roughly at its geometry center. Panel (e) shows the variation of $T_{g}$ of selected field lines along the distance from its axis. Panels (f) and (g) display the side and top views of the distributions of $Q$ factor and the magnetic reconnection configuration at 09:00~UT on 2017 September 6, respectively. The dark-cyan tubes represent the pre-reconnection flux rope field lines, pink tubes show the pre-reconnection sheared arcades, and yellow tubes depict the post-reconnection field lines. The 3D semi-transparent surfaces represent a high $Q$ value of lg $Q>$ 3.}
\label{figure2}
\end{figure}

The manifestation of the effects of the photospheric motions on coronal magnetic fields is well represented by the temporal evolution of the field lines. At 19:00 UT on September 3, the field lines are nearly potential and perpendicular to the PILs (Figure~\ref{figure1}a). After 30 hours, a twisted flux rope and some peripheral sheared arcades are formed above and alongside the central PIL (Figure~\ref{figure1}c). The upper panels of Figure~\ref{figure2} provide an overview of the formed flux rope shown with representative field lines (Figures~\ref{figure2}a and \ref{figure2}d), the squashing factor $Q$ (Figure~\ref{figure2}b), and the twist number (Figures~\ref{figure2}c and \ref{figure2}e). The $Q$ factor is computed using an open-source code (K-QSL) developed by Kai E.\ Yang\footnote{https://github.com/Kai-E-Yang/QSL}. Quasi-separatrix layers (QSLs) depict regions where magnetic connectivity changes drastically \citep{Priest1995, demo1996} and are often used to delineate flux-rope boundaries \citep{Titov1999, Aulanier2010, Janvier2013, Dudic2014, Guo2013, Guo2017, Aulanier2019, Guojh2021}. Two types of twist number, denoted as $T_{w}$ and $T_{g}$, are computed to quantify the degree of twist of the flux rope, as shown in Figures~\ref{figure2}c and \ref{figure2}e, which are described as follows \citep{Berger2006}:
\begin{eqnarray}
&& T_{w} = \int \frac{\mu_{0}J_{\parallel}}{4\pi B}dl,\label{eq10}\\
&& T_{g} = \frac{1}{2\pi} \int \boldsymbol{T}(s)\cdot \boldsymbol{V}(s) \times \frac{d\boldsymbol{V}(s)}{ds}ds, \label{eq11}
\end{eqnarray}
where $\boldsymbol{V}(s)$ denotes the unit vector normal to $\boldsymbol{T}$(s) and pointing from the flux-rope axis to the other curve. $T_{w}$ is calculated with the parallel electric current using the code implemented by \citet{Liu2016}, representing the twist degree between two infinitesimally close field lines. In contrast, $T_{g}$ is directly computed from the geometry of the field lines \citep{Berger2006}, which can describe how many turns of the field lines wind about one common axis. An open-source code to compute $T_{g}$ can be found on GitHub\footnote{https://github.com/njuguoyang/magnetic\textunderscore modeling\textunderscore codes}. More details and comparisons between the two metrics can be found in \citet{Liu2016} and \citet{Price2022}. Following our previous works \citep{Guo2013, Guo2017, Guojh2021}, the flux-rope axis is taken as a nearly non-twisted field line roughly passing through the geometry center (red dot in Figure~\ref{figure2}c), and the boundary is determined by $Q$ and $T_{w}$ maps. As shown in Figures~\ref{figure2}a and \ref{figure2}d, one can see that the field lines are twisted with strong electric current inside, and the $Q$-map displays a quasi-circular shape enveloping the boundary of the flux rope. Moreover, certain field lines exhibit the twists exceeding one turn, both in terms of the $T_{w}$ and $T_{g}$ metrics. The analysis of the magnetic topology strongly indicates the formation of a twisted flux rope during the long-term evolution of the active region.

Thereafter, it is found that some flux-rope field lines connect to the northern remote negative polarities (Figure~\ref{figure1}e), indicating the growth of the flux rope. To decipher this, we calculate the 3D distribution of $Q$ factor and plot some representative field lines around the QSLs in Figures~\ref{figure2}f and \ref{figure2}g. One can see X-shaped QSLs, distinguishing the field lines with different connectivities very well. The field lines in the left and right parts of the X-shaped structure are northern sheared arcades (pink lines) and twisted flux-rope field lines (dark-blue lines), respectively. However, the field lines in the top part connect to the remote negative polarities in the north (yellow lines). Consequently, there should exist magnetic reconnection between the flux rope (dark-blue lines) and sheared arcades (pink lines), which leads to the growth of the flux rope. This topological structure is also found in the simulations conducted by \citet{Inoue2021} and \citet{Price2019}. Apart from that, a null-point structure (pink lines in Figures~\ref{figure1}e and \ref{figure1}f) is found on one side of the twisted flux rope. To compare our simulation results with observations, we overlay some sample field lines on observed 304~\AA\ images in Figures~\ref{figure1}b, \ref{figure1}d, and \ref{figure1}f. It is found that the bright filamentary loops in 304 \AA\ observations resemble the magnetic loops in the simulation (Figure~\ref{figure1}b), and the overall shape of the filament is similar to the flux rope (Figure~\ref{figure1}f). 

\subsubsection{Quantitative evolution of the magnetic energy and relative magnetic helicity} 

\begin{figure}[ht!]
\includegraphics[scale=0.4]{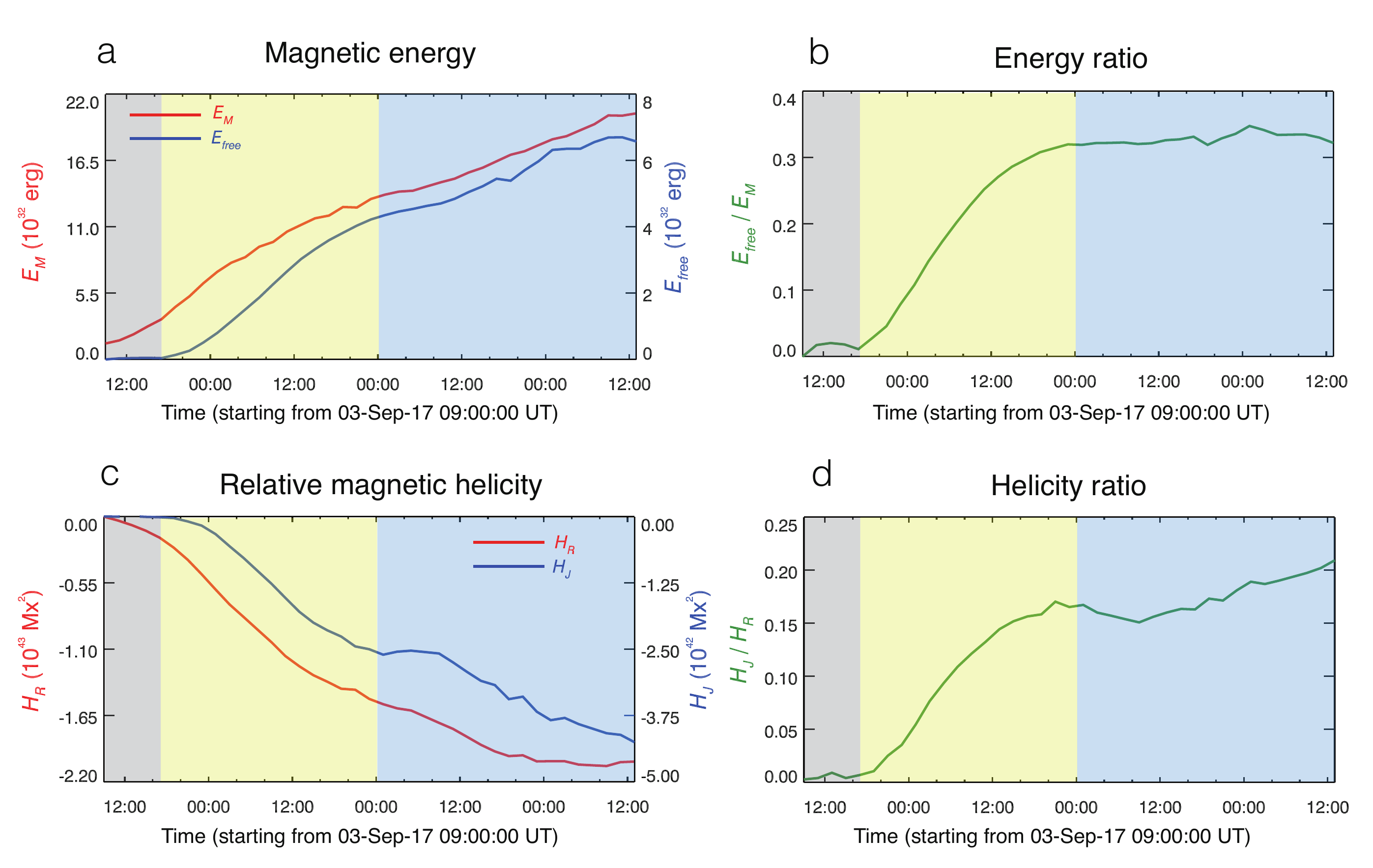}
\centering
\caption{Temporal evolution of the magnetic energy (a), the ratio of the free energy to total magnetic energy (b), relative magnetic helicity (c), and the ratio of the current-carrying helicity and the total relative helicity (d). The red and blue lines in panel (a)/(b) denote the evolution of the total and free magnetic energy/the total relative helicity and current-carrying helicity, respectively. The gray, yellow, and blue bands in each panel correspond to slow rise, rapid injection, and gradual phases, respectively.}
\label{figure3}
\end{figure}

To further elaborate on the accumulation of magnetic energy, we compute both the total magnetic energy $E_{M}$ and the free magnetic energy $E_{free}$ within the simulation domain as:
\begin{eqnarray}
  && E_{M}=\frac{1}{2\mu_{0}}\int \boldsymbol{B}^{2} dV,\label{eq12}\\
  && E_{free}=\frac{1}{2\mu_{0}}\int (\boldsymbol{B}^{2}-\boldsymbol{B_{p}}^{2}) dV,\label{eq13}\\
 \notag  
 \end{eqnarray}
where $\boldsymbol{B}$ represents the total magnetic field, $\boldsymbol{B_{p}}$ corresponds to the potential field, and $\mu_{0}$ is the permeability of free space. Figure~\ref{figure3}a illustrates the temporal evolution of the total magnetic energy (red curve) and free energy (blue curve), and Figure~\ref{figure3}b shows their ratio ($E_{free}/E_{M}$). It is found that the total magnetic energy ($E_{M}$) displays a monotonic increase over time, while the free energy ($E_{free}$) and the ratio ($E_{free}/E_{M}$) manifest a more intricate behavior that can be roughly divided into three distinct stages. During the initial stage (grey band), even though the total magnetic energy almost exhibits a continual increase, the free energy and the ratio remain relatively stable at low values. This suggests that there is no substantial accumulation of nonpotential energy during this period. In the subsequent phase (yellow band), both the free energy and ratio undergo a pronounced increase, with the ratio escalating to 0.32. Following this, despite the continued elevation of both total and free magnetic energy, a plateau is encountered in their ratio, wherein the rate of increase begins to level off, as depicted by the blue band.

Magnetic helicity is an effective scalar quantity to characterize the topological complexity of the magnetic field \citep{Valori2016}. Therefore, to unveil the evolution of the complexity of the field-line connectivity, we compute the relative magnetic helicity of the entire simulation domain. In practice, the gauge-independent relative helicity with respect to the reference potential field (the condition of $\boldsymbol{\hat{n} \cdot B= \hat{n} \cdot B_{p}}$ is satisfied on the boundary), which can be described by the following equations \citep{Berger1984, Berger1999}:
 \begin{eqnarray}
  && H_{R}=\int (\boldsymbol{A}+\boldsymbol{A_{p}})\cdot(\boldsymbol{B}-\boldsymbol{B_{p}}) dV=H_{J}+H_{pJ},\label{eq14}\\
    && H_{J}=\int (\boldsymbol{A}-\boldsymbol{A_{p}})\cdot(\boldsymbol{B}-\boldsymbol{B_{p}}) dV,\label{eq15}\\
  && H_{PJ}=2\int \boldsymbol{A_{p}}\cdot(\boldsymbol{B}-\boldsymbol{B_{p}}) dV,\label{eq16}\\
 \notag  
 \end{eqnarray}
where $\boldsymbol{A}$ represents the vector potential of the total magnetic field $\boldsymbol{B}$, $\boldsymbol{A_{p}}$ denotes the vector potential of the corresponding potential magnetic field $\boldsymbol{B_{p}}$, $H_{R}$ is the total magnetic helicity, $H_{J}$ is the helicity of the current-carrying part, and $H_{P,J}$ is the volume-threading helicity between the potential field and the current-carrying field. The computation is executed using the method based on the DeVore gauge \citep{Devore2000a, Devore2000b, Valori2012} and realized in \citet{Yu2023}, namely, $A_{z}=A_{p,z}=0$, in which the vector potential $A$ can be explicitly described with the straightforward integration of the magnetic fields. 

The evolution of total helicity ($H_{R}$) and current-carrying helicity ($H_{J}$) is depicted by the red and blue curves in Figure~\ref{figure3}c, respectively, and the helicity ratio ($H_{J}/H_{R}$) is illustrated in Figure~\ref{figure3}d. It is found that the helicity evolves similarly to magnetic energy. The total helicity ($H_{R}$) roughly increases in magnitude throughout the simulation. The curves of the current-carrying helicity ($H_{J}$) and the helicity ratio ($H_{J}/H_{R}$) also exhibit three distinct stages:~slow rise, rapid injection, and gradual phases. Additionally, it is seen that the helicity ratio exhibits more pronounced fluctuations compared to the energy ratio, which may arise from the inherent ability of helicity to unveil intricate details of the magnetic system. Our simulation results substantiate the findings of previous works \citep{Phillips2005, Pariat2017, Price2019, Linan2018, Linan2020}, highlighting that free energy, current-carrying helicity, and energy/helicity ratio are more effective in discriminating the distinct stages of active region evolution compared to the total energy and helicity. 

\subsubsection{Formation mechanism of the magnetic flux rope} 

\begin{figure}[ht!]
\includegraphics[scale=0.83]{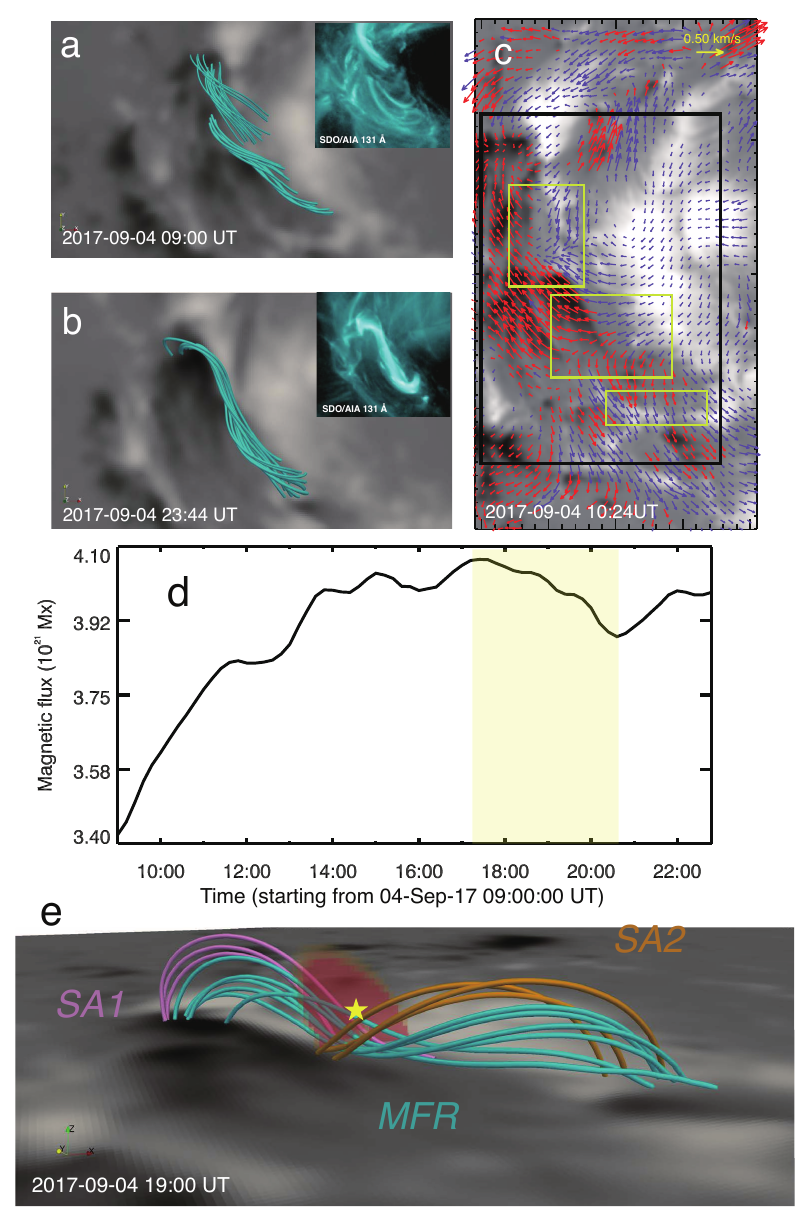}
\centering
\caption{Formation process of the flux rope in the TMF model. Panels (a) and (b) show the sheared arcades and flux ropes, respectively, where the insets show the SDO/AIA 131 \AA\ images at the same time. Panel (c) shows the horizontal velocity fields in the photosphere, where the yellow rectangles delineate the regions where collisional shearing could potentially occur, and the black rectangle represents the area used for the magnetic flux calculation. Panel (d) displays the temporal evolution of the magnetic flux of negative polarity computed from the area marked by the black rectangle in panel (c). The yellow band corresponds to the time intervals during which the flux cancellation unfolds. Panel (e) illustrates the reconnection configuration of the flux rope, including two groups of sheared arcades prior to reconnection (SA1, SA2), and newly reconnected flux rope, which is wrapped by the current sheet depicted by the high $J/B$ region (magenta region). The pentagram represents the reconnection site.}
\label{figure4}
\end{figure}

Figure~\ref{figure3} illustrates a significant increase in the free magnetic energy and current-carrying helicity on 2017 September 4, implying that the twisted flux rope is likely formed during this period. To demonstrate this development, we present some typical field lines above the PILs at the beginning and end of this period in Figures~\ref{figure4}a and \ref{figure4}b. At 09:00 UT on September 4, we find two sets of sheared arcades that are reminiscent of the bright sheared loops seen in AIA 131 \AA\ observations, and no twisted field lines can be identified. However, at 23:44 UT, the sheared arcades evolve into a twisted flux rope, resembling the sigmoid-shaped hot channel observed at the same time. This noteworthy transformation strongly suggests that the flux rope formation takes place during this period.

To explore the underlying mechanisms of the flux rope formation, we present the distribution of the photospheric flows at 10:24 UT in Figure~\ref{figure4}c, which reveals the presence of strong shearing flows around the central PIL. Additionally, we identify converging motions where the positive polarities collide with the negative ones (marked by yellow rectangles). These characteristic photospheric flows may correspond to what is termed ``collisional shearing" \citep{Chintzoglou2019}, which is commonly invoked to explain the flux rope formation within complicated active regions including multiple emerging bipoles. Particularly, \citet{Liu2019} demonstrated the effectiveness of collisional shearing to explain the flux rope formation in this active region, based on observations and NLFFF extrapolations.

As described in \citet{Chintzoglou2019}, one of the key outcomes of collisional shearing is flux cancellation. This process is widely believed to be the pivot in initiating the formation of a twisted flux rope \citep{Van1989}. To investigate this, we present in Figure~\ref{figure4}d the temporal evolution of the unsigned vertical flux due to the negative polarity within the black rectangle shown in Figure~\ref{figure4}c. Evidently, the reduction of about 5$\%$ can be seen (yellow band), indicating flux cancellation. Particularly, we do not account for the fluxes entering and leaving the computation box, which means that not all of the reduction is solely attributed to flux cancellation. Additional observational evidence of flux cancellation in this active region can be found in \citet{Liu2019}. Figure~\ref{figure4}e displays the magnetic configuration at 19:00~UT. We can identify a high $J/B$ region, where the associated current sheet is color-coded as magenta and the field lines around are shown as purple and cyan lines. The current sheet displays an X-shaped configuration, with two J-shaped arcades and a twisted flux rope. This result is in agreement with the flux cancellation model to explain the formation of a flux rope \citep{Van1989}. Combing these results, our data-driven modeling suggests that the collisional shearing, along with the resulting flux cancellation, leads to the formation of the flux rope.

\subsection{Drastic evolution during the X2.2 confined flare}\label{ce}

\subsubsection{Global evolution and comparison with observations}

\begin{figure}[ht!]
\centering
\includegraphics[scale=0.13]{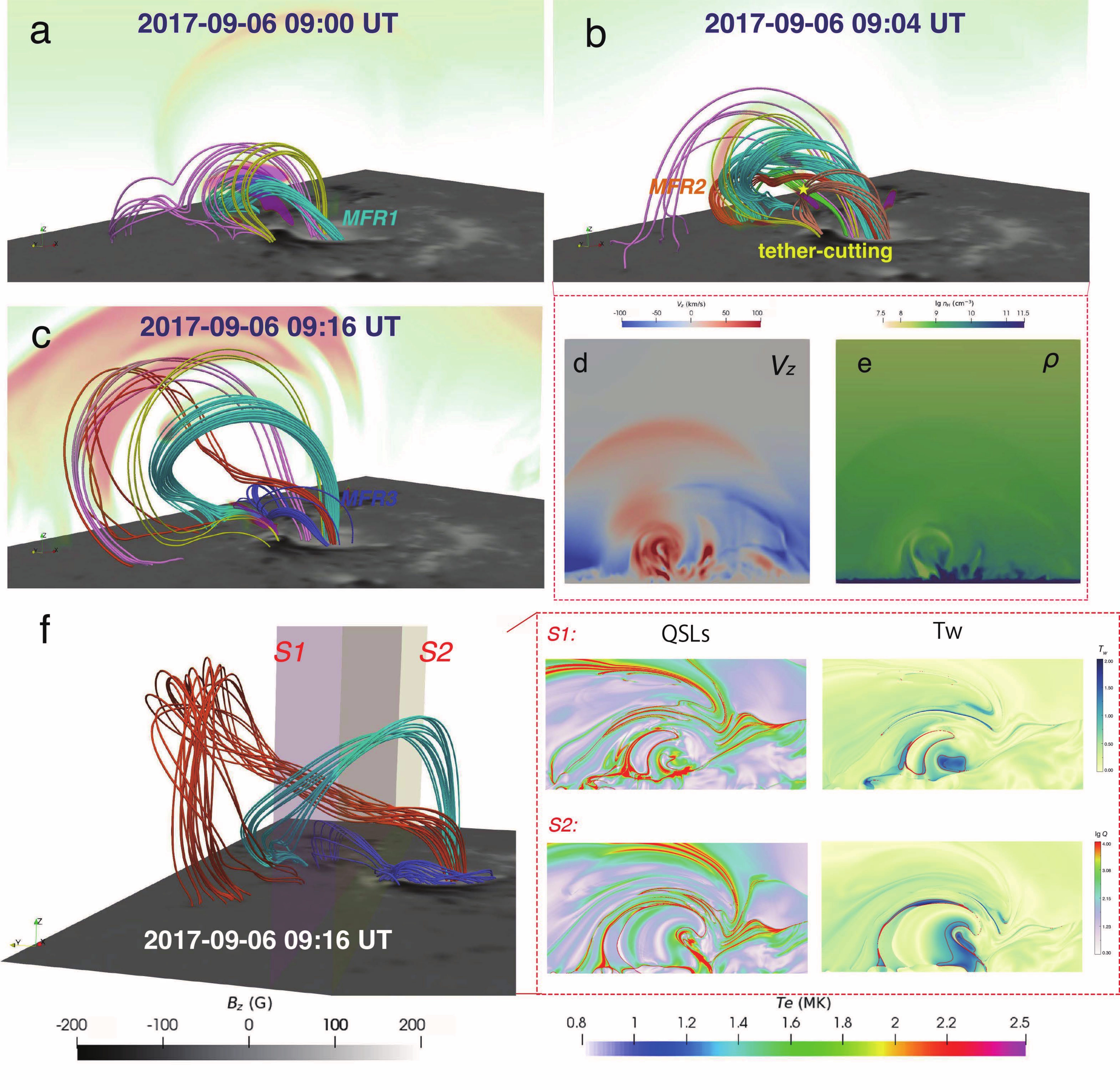}
\caption{Magnetic field lines and temperature distributions at 09:00 UT (a), 09:04 UT (b), and 09:16 UT (c) on 2017 September 6. The cyan, orange, and blue lines represent the original flux rope (MFR1) before eruption, and two newly formed ones (MFR2 and MFR3) during eruption. The pentagrams depict the reconnection sites. Panels (d) and (e) show the distributions of $V_{z}$ and $\rho$ at 09:04~UT on the plane showing temperature. Panel (f) illustrates three flux ropes at the end of simulation. The accompanying figures on the right side show the maps of squashing factor $Q$ and twist number $T_{w}$ in two vertical semi-transparent surfaces. The red contours in $T_{w}$ maps delineate the values of $T_{w}=$1, 1.5 and 2.}
\label{figure5}
\end{figure}

\begin{figure}[ht!]
\centering
\includegraphics[scale=0.8]{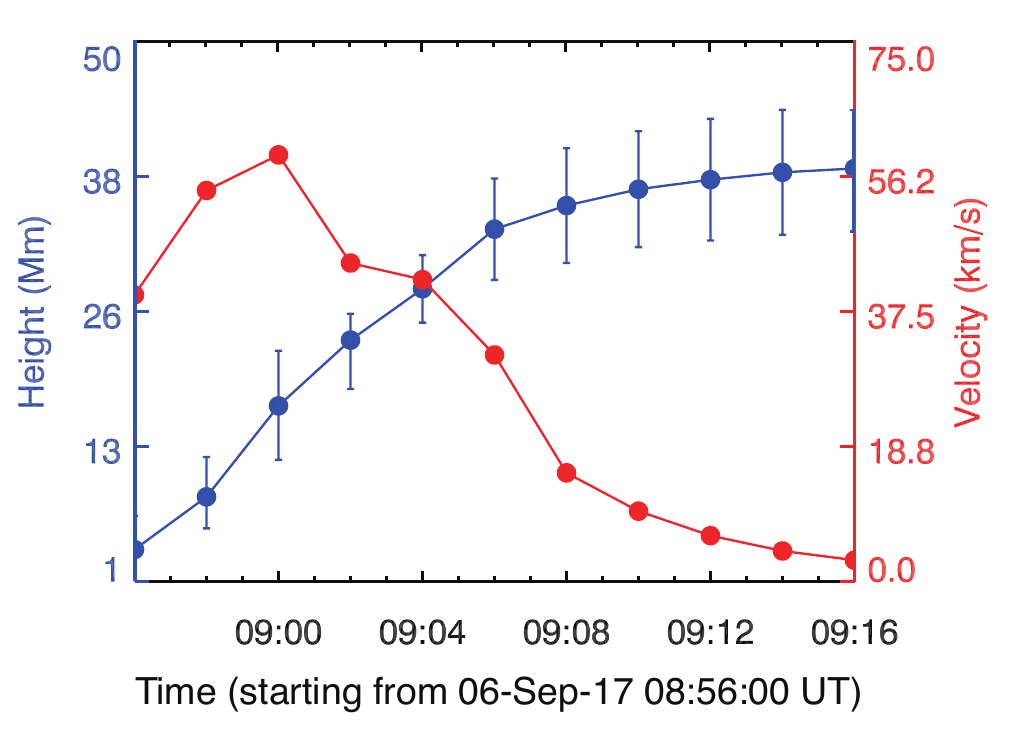}
\caption{Time–distance measurement of the flux rope in the MHD simulation. The blue and red dots show the height and velocity, respectively. The error bars of the height are determined by the uncertainty with repeated measurements of ten times.}
\label{figure6}
\end{figure}

Figure~\ref{figure5} displays the dynamic evolution of the magnetic configuration during the eruption simulated by the  thermodynamic MHD model. It is seen that the flux rope is not a typical coherent one. It includes three  distinct branches, i.e., MFR1 (cyan lines), MFR2 (orange lines), and MFR3 (blue lines). They are roughly discernible through the heated volumes or field-line connectivity proxies, including the $Q$ and $T_w$ maps. As the flux rope MFR1 rises, drastic tether-cutting reconnection in the current sheet underneath the flux rope is induced. Double J-shaped arcades (brown and green lines) reconnect to form a new flux rope, labeled as MFR2, as illustrated in Figure~\ref{figure5}b, which means that the core structure of this eruption event is comprised of multiple flux ropes. We find that the southern footpoint of MFR1 moves northward, which could be due to magnetic reconnection between the flux rope and the ambient arcades \citep{Aulanier2019}. The temperature around the reconnection site is enhanced to about 3~MK. Figures~\ref{figure5}d and \ref{figure5}e display the distributions of vertical velocity and density, respectively, from which we can see the eruptive flux rope and its driven shock. Subsequently, MFR1 starts to rotate counterclockwise until its top part is almost parallel to the overlying potential field lines. Eventually, MFR1 rises slowly and almost halts at the height of 40~Mm. Figure~\ref{figure6} showcases the kinematics of MFR1. It is evident that the flux rope experiences gradual deceleration after initial impulsive acceleration. By 09:16~UT, the speed of the flux rope decreases to a value of less 3~km s$^{-1}$, suggesting that the eruption simulated by our MHD model is a confined one. It is worth noting that, at 09:16 UT, a third flux rope, labeled as MFR3, is formed at a lower height inside the flaring loops (Figure~\ref{figure5}c). It is noted that all the three flux ropes have negative helicity, which is in line with that of the active region.

Moreover, to further clarify the topological properties of different flux-rope branches, we calculate the squashing factor $Q$ distributions at two vertical planes, Slices 1 and 2, which are shown as S1 and S2 in Figure~\ref{figure5}f. We find that MFR1 and MFR2 shown in Slice 1 display separate closed QSLs, implying that they are indeed individual coherent flux ropes. On the other hand, MFR3 is less coherent and presents only a hyperbolic flux tube (HFT) configuration in S2, suggesting it is still in the formation process. It should be emphasized that the QSLs of MFR1 in S2 are different from those in S1, which indicates that the flux rope in observations is far from being translational invariant.

\begin{figure}[ht!]
\centering
\includegraphics[scale=1.1]{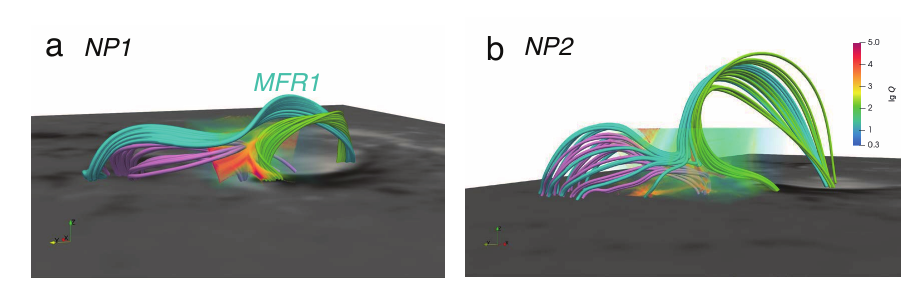}
\caption{Two null-point structures around the flux rope at 08:58~UT on 2017 September 6. Panel (a) highlights the first null point, NP1, and panel (b) highlights the second one, NP2. The green and pink tubes represent the pre-reconnection field lines, and cyan tubes depict the post-reconnection field lines. The 3D semi-transparent surfaces visualize the distributions of $Q$.}
\label{figure7}
\end{figure}

\begin{figure}[ht!]
\centering
\includegraphics[scale=0.68]{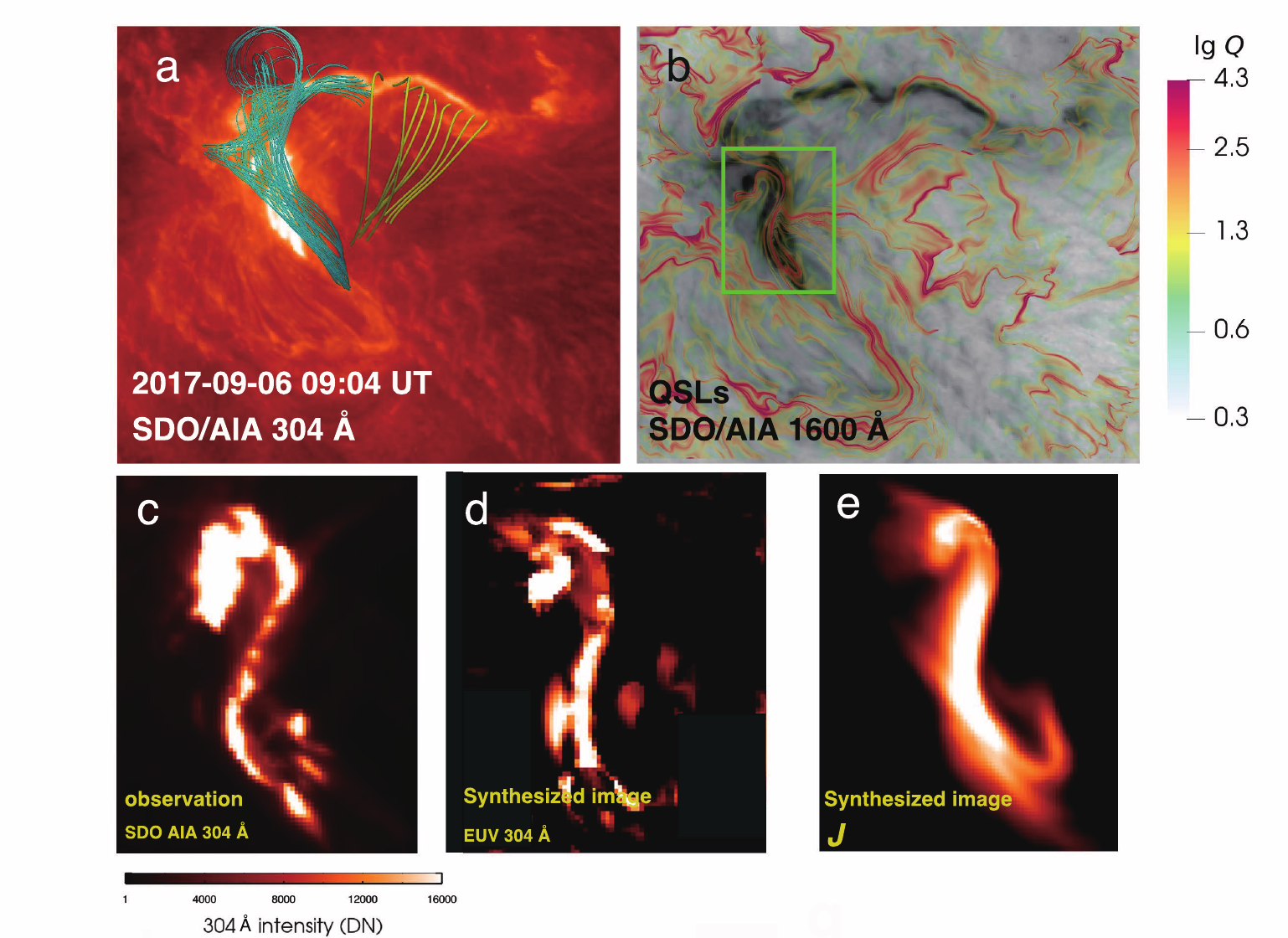}
\caption{Comparison between simulations and SDO/AIA observations at 09:04 UT on 2017 September 6. (a) The 3D eruptive structure overlaid on an AIA 304 \AA\ image. (b) QSL distribution at the solar surface overlaid on an AIA 1600 \AA\ image. (c) Zoomed-in SDO/AIA 304 \AA\ image of the green rectangle region in panel (b). (d) Synthesized EUV 304 \AA\ radiation image. (e) Synthesized image computed by the integration of the electric current density along the vertical direction. }
\label{figure8}
\end{figure}

To comprehend the initiation of the X2.2 flare, we examine the magnetic topology at the flare onset (08:58~UT), where we can identify two null-point reconnection sites encircled by QSLs, as depicted in Figure~\ref{figure7}. This two null-point structure was also identified in previous numerical modelings of this active region \citep{Price2019, Inoue2021, Daei2023}. The first nullpoint, NP1, is positioned between the flux rope (green lines) and ambient arcades (pink lines), resulting in a reconstruction of the original flux rope, and the formation of MFR1 in Figure~\ref{figure5}. This is also demonstrated to be crucial in the formation and rising of the flux rope, as shown in Figure 6 of \citet{Daei2023}. Besides, this reconnection geometry can be classified as the ar-rf (arcade-rope to rope-flare loop) reconnection in the 3D flare model \citep{Aulanier2019, Dudik2019}. Furthermore, this null-point configuration acts as a linkage between the northern negative polarity group and the central PILs, playing a crucial role in forming the northern remote flare ribbons shown in Figures~\ref{figure8}a and \ref{figure8}b. The second nullpoint, NP2, is located alongside the flux rope and appears in the overlying background fields. The magnetic reconnection in both types of null-point structures can initiate the ascent of the flux rope, as demonstrated by \citet{Chen2000}. These two topological structures have also been found and demonstrated to be closely associated with the eruption in previous works \citep{Mitra2018, Price2019, Bamba2020, Zou2020, Inoue2021, Yamasaki2021}.

Figure~\ref{figure8} exhibits a comparison between the simulation results and observations. First, in Figure~\ref{figure8}a, we display some typical field lines of the eruptive structure overlaid on an AIA 304 \AA\ image. It is seen that the magnetic connectivity resembles the main flare ribbons and a remote ribbon very well. Second, we overplot the simulated QSLs at the solar surface on the flare ribbons observed by AIA 1600 \AA\ waveband in Figure~\ref{figure8}b. It is evident that the QSLs in the model align well with the flare ribbons, especially the core inverse S-shaped structure. However, the northwest ribbon linked by the yellow lines in Figure~\ref{figure8}a is not reproduced very well. Nevertheless, the above results indicate that the key magnetic topological structure of this eruption event is almost captured by our simulation with minor exceptions. We then compare the synthesized emission derived from the simulation (Figures~\ref{figure8}d and \ref{figure8}e) with an AIA 304 \AA\ image in observations (Figure~\ref{figure8}c). Among this, Figure~\ref{figure8}d shows more realistic radiation computed from the temperature and density in the simulation \citep{Guo2023}, while Figure~\ref{figure8}e is a mock radiation image obtained by integrating the electric current density along the $z$-axis \citep{Zhong2021}. Both synthesized images successfully reproduce the inverse S-shaped structure in the observations, while the synthesized electric-current image appears to be too smooth compared to the observation and synthesized radiation image. In summary, our data-driven model provides a good match to the observations, including the dynamics of the flux rope, magnetic topology, and emission. These results demonstrate the reliability of our simulation.

\subsubsection{Confining mechanism according to the decomposition of Lorentz force }

Our numerical model successfully reproduced the observed confined eruption, providing us with an opportunity to explore which physical mechanisms restrain the eruption from being successful. To investigate this quantitatively, we adopt the Lorentz force decomposition as done in \citet{Zhong2021}, where the key Lorentz force contributing to a solar eruption can be decomposed into three terms \citep{Myers2015}: the strapping force ($F_{_{S}}$), the hoop force ($F_{_{H}}$), and the tension force ($F_{_{T}}$). Among them, the strapping force and the hoop force originate from the poloidal magnetic fields, while the tension force results from the toroidal magnetic fields, as described in \citet{Zhong2021}. 

\begin{figure}[ht!]
\centering
\includegraphics[scale=0.9]{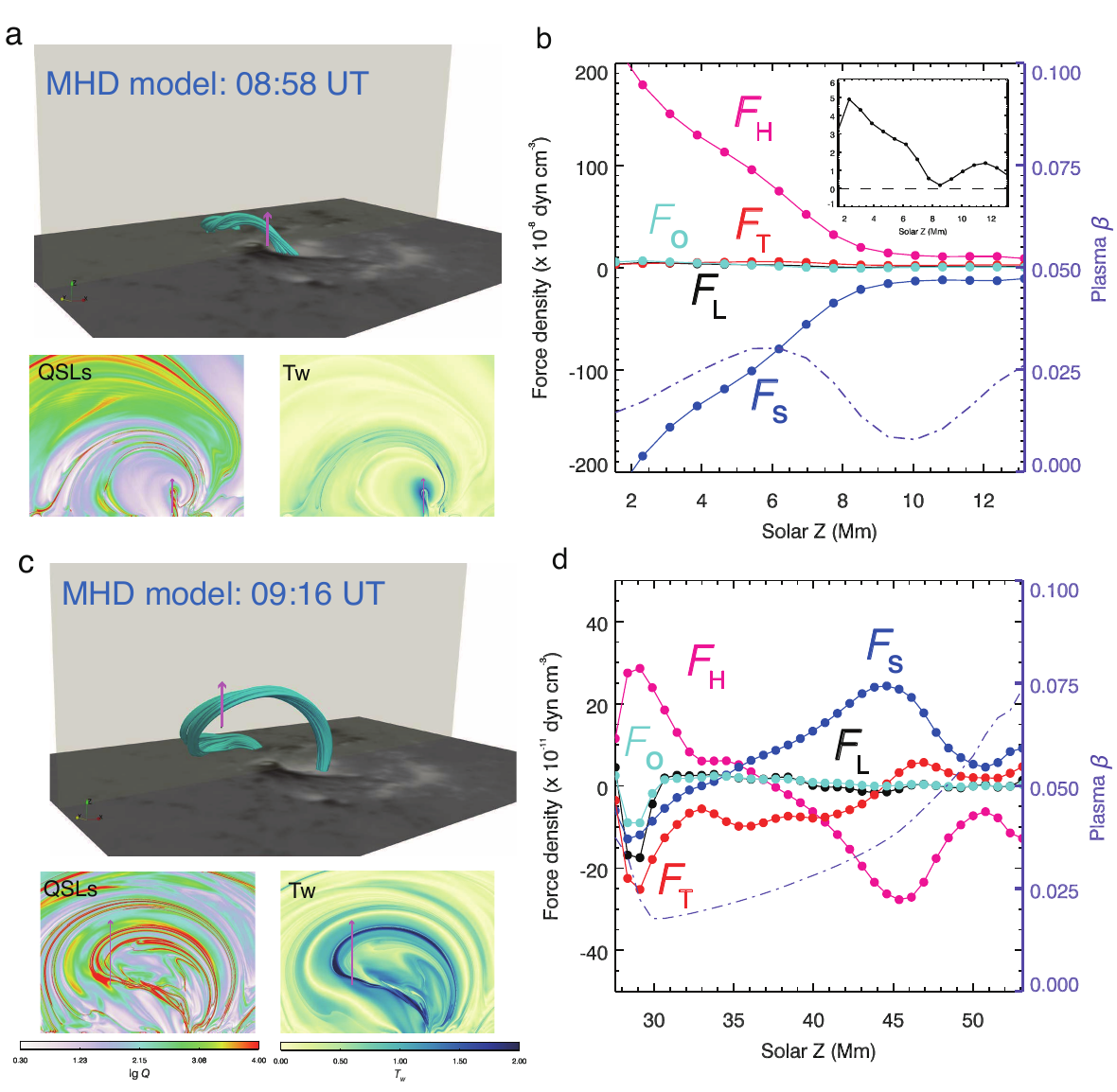}
\caption{Comparison between the flux-rope structures and the vertical Lorentz force. (a, c) 3D illustrations of the magnetic configuration of MFR1, and distributions of $Q$ factor and $T_{w}$ (in the gray semi-transparent surfaces) at 08:58 UT (a) and 09:16 UT (c) on 2017 September 6. The pink arrows extend from the bottom to the top of the flux rope. (b, d) Distributions of the vertical component of the Lorentz force along the pink arrows in panels (a) and (c), where the black, pink, blue, red, and cyan lines linked by circles represent the net Lorentz force ($F_{_{L}}$), hoop force ($F_{_{H}}$), strapping force ($F_{_{S}}$), tension force ($F_{_{T}}$), and residual force ($F_{_{O}}$), respectively. The purple dashed lines represent the distributions of plasma $\beta$ along pink arrows.}
\label{figure9}
\end{figure}

Figures~\ref{figure9}a and \ref{figure9}c show the magnetic configurations of MFR1 and the corresponding $Q$ and $T_{w}$ maps at two moments, where the $Q$ and $T_w$ maps delineate the boundaries of the flux ropes. Figures~\ref{figure9}b and \ref{figure9}d show the spatial distributions of different Lorentz force components along two pink arrows across the flux rope at 09:00 UT and 09:16~UT, respectively. We find that the net Lorentz force ($F_{_{L}}$) is significantly smaller in magnitude compared to its components,  particularly the hoop force ($F_{_{H}}$) and strapping force ($F_{_{S}}$). Nevertheless, the net Lorentz force consistently maintains an upward direction along the designated axis at 09:00~UT (shown in the insert panel of Figure~\ref{figure9}b), thereby leading to the rising of the flux rope. However, the net Lorentz force is significantly downward in the $z$ range from 28 to 31~Mm at 09:16~UT (Figure~\ref{figure9}d), constraining the further rising-up of the flux rope. In particular, the downward force is mainly contributed by the tension force instead of the strapping force, which is different from the initial state. This scenario can be understood as follows: the tension force originates from the toroidal magnetic field (the axial magnetic field of the flux rope), so when the flux rope rotates during the eruption, the orientation of the external magnetic field would transfer from the poloidal to toroidal direction, decreasing the strapping force but significantly increasing the tension force. Furthermore, to evaluate if the magnetic field dominates over the gas pressure in this case, we plot the distributions of plasma $\beta$ (purple dashed lines in Figures~\ref{figure9}b and \ref{figure9}d). It reveals that the plasma $\beta$ value is below 0.1, indicating that the confined eruption is magnetically dominated.

\section{Discussions}\label{sec:discussion}

\subsection{Why do some torus-unstable flux ropes fail to erupt?}\label{cm}

One interesting aspect of CME research is that not all rising flux rope eruptions can eventually escape into interplanetary space and evolve into ICMEs. In fact, some flux ropes are confined in the solar atmosphere. For example, \citet{Ji2003} observed the first failed eruptions by tracing the motions of filament materials. They found that the filament starts to rise with evident rotation, reaches a maximum height, and then falls back to the solar surface. To explain such a phenomenon, \citet{Torok2005} performed a data-inspired simulation and suggested that the initiation of the filament eruption and the rotation was triggered by the kink instability. Regarding the confining mechanism, it is well accepted that the decrease of the overlying magnetic field with height, i.e., the torus instability \citep{Kliem2006}, determines whether the flux rope eruption will succeed or fail. It was claimed that the decay index of the background magnetic field has a threshold of 1.5, above which the situation leads to a successful CME and below which the situation leads to a failed eruption even when the flux rope satisfies the kink instability criterion. However, it has been noticed that such a threshold was derived with the condition that the flux rope satisfies the toroidal symmetry assumption and is slender enough, which is representative of the infinite aspect ratio of the flux rope. Under this assumption, a remarkably concise instability criterion of $n>1.5$ is derived in the circuit framework (not the MHD framework), and the instability is mainly dominated by the hoop force and strapping force, corresponding to $c_{0} \to \infty$ in Equation~(5) of \citet{Kliem2006}. That being said, this $n_{\rm crit}=1.5$ instability criterion is primarily responsible for the cases that are suppressed by the downward strapping force \citep{Myers2016}.

Recently, \citet{Zhou2019} conducted a statistical study of 16 failed filament eruptions, and examined the relevance between the decay index of the overlying fields and the rotation angle during eruption. Strikingly, they found that all the torus-unstable events ($n > n_{crit}=1.5$) displayed large-angle rotations (exceeding $40^{\circ}$), suggesting that there should exist some physical connection between confined eruptions and the flux rope rotation. Since the magnetic tension force results from the toroidal magnetic field, its effects are supposedly magnified in the flux rope rotation cases. After the rotation, the flux rope axis may become more parallel to the overlying fields, which would cause a direction change of the background field from being poloidal to being toroidal. As a result, the strapping force caused by the background poloidal magnetic field would decrease, while the tension force originating from the toroidal magnetic field would increase significantly, as shown in Figure~\ref{figure9}d. In this scenario, the toroidal-field tension force, rather than the poloidal-field strapping force, becomes the primary constraining force accounting for the confined eruption. Therefore, based on our data-driven model, we suggest that the confining mechanism for the rotation events might be mainly attributed to the tension force instead of the strapping force. This result is consistent with the findings of laboratory plasma experiments \citep{Myers2015}, who find that confined eruptions in the failed torus regime is dominated by the dynamic tension force. In addition to that, the effects of tension force on confining solar eruptions are also reported by \citet{Joshi2022} and \citet{Wang2023}. Recently, a similar scenario was also presented in the MHD simulation based on an ideal bipolar configuration performed by \citet{Jiang2023}, who found that the tension force can halt the rising of the flux rope accompanied by rotation even though the value of the decay index has exceeded 1.5.

Our data-driven simulation self-consistently explains why many rotation events fail to be eruptive although they are torus-unstable ($n>1.5$). Our results also imply that it is important to exercise caution when calculating the decay index criterion, i.e., it might be necessary to measure the temporal evolution of the toroidal and poloidal directions as the flux rope ascends. In addition, several studies have found that there is a variation for the torus instability criterion of $n>1.5$. For instance, \citet{Demoulin2010} found that the decay index for a relatively thick current can decrease to a value of 1.1. \citet{Zuccarello2015} conducted a series of MHD simulations and identified a critical range for the decay index between 1.3 and 1.5, challenging the universal critical value of 1.5 for the onset of torus instability. Besides, \citet{Myers2015} found that the torus instability criterion can decrease to a value of 0.8 in laboratory experiments. Moreover, even if the background magnetic field satisfies the torus instability criterion at any height, it does not guarantee the eruption will be successful otherwise it would disobey the Aly-Sturrock constraint \citep{Chen2020}. Consequently, it is a caveat to simply employ the decay index value of 1.5 to determine the eruption. In this paper, we adopt the perspective of Lorentz force decomposition to investigate the confining mechanism of flux rope eruptions, providing an alternative perspective for exploring the underlying physical processes of solar eruptions.
 
\subsection{Rapid buildup of twisted flux ropes during the confined eruption}\label{rb}

The occurrence of homologous and successive flares inside the same active region within hours is a frequently observed phenomenon \citep{Yang2017}. In particular, confined flares are sometimes observed prior to eruptive flares. This happens in our case of active region 12673 where a major X9.3 flare along with a CME is observed three hours after the confined X2.2 flare. It is intriguing to explore how magnetic energy is accumulated within hours after much of the magnetic energy has already been released in the primary flare. Some researchers claimed that during the confined eruption, a flux rope is formed via magnetic reconnection in the primary flare, which can facilitate the successful eruption of the subsequent eruptive flares. For example, \citet{Guo2013} reported the growth of an eruptive flux rope during a series of confined eruptions. \citet{Patsourakos2013} observed the formation of a flux rope during a confined flare, which subsequently evolves into a CME. \citet{Liu2018} found the rapid build-up of magnetic helicity and axial fluxes of a flux rope during a confined X2.2-class flare. Although these studies have pointed out that magnetic free energy can accumulate during confined eruptions based on NLFFF magnetic extrapolations of sequential magnetograms on the solar surface, the dynamic evolution of this process has not yet been well-reproduced by data-driven simulations.

In this paper, we found that only one flux rope, i.e., MFR1, exists before the X2.2-class flare. During the flare, a second flux rope is formed due to the flare-associated magnetic reconnection. Generally, it is believed in the classical standard CME/flare model \citep{Chen2011}, that the newly reconnected flux joins the pre-existing flux rope, simply increasing the poloidal flux of the flux rope. However, our simulation results indicate that the newly reconnected flux might form a separate flux rope, i.e., MFR2. This is similar to the formation of multiple plasmoids in 2D MHD simulations of magnetic reconnection. The two flux ropes do not coalesce. Instead, there exist a QSL between them, as shown in Figure~\ref{figure5}f. More interestingly, it is revealed that a third flux rope, MFR3, is being formed inside the flaring loops due to the continual shearing and converging flows on the solar surface. The formation of MFR3 due to the low-atmosphere magnetic reconnection is evidenced by the bald patch and the low-lying HFT structures around MFR3. In particular, all three flux ropes posses negative helicity. We propose that MFR3 is likely to be the core structure of the subsequent successful eruption associated with the ensuing X9.3-class flare. 

\begin{figure}[ht!]
\centering
\includegraphics[scale=0.5]{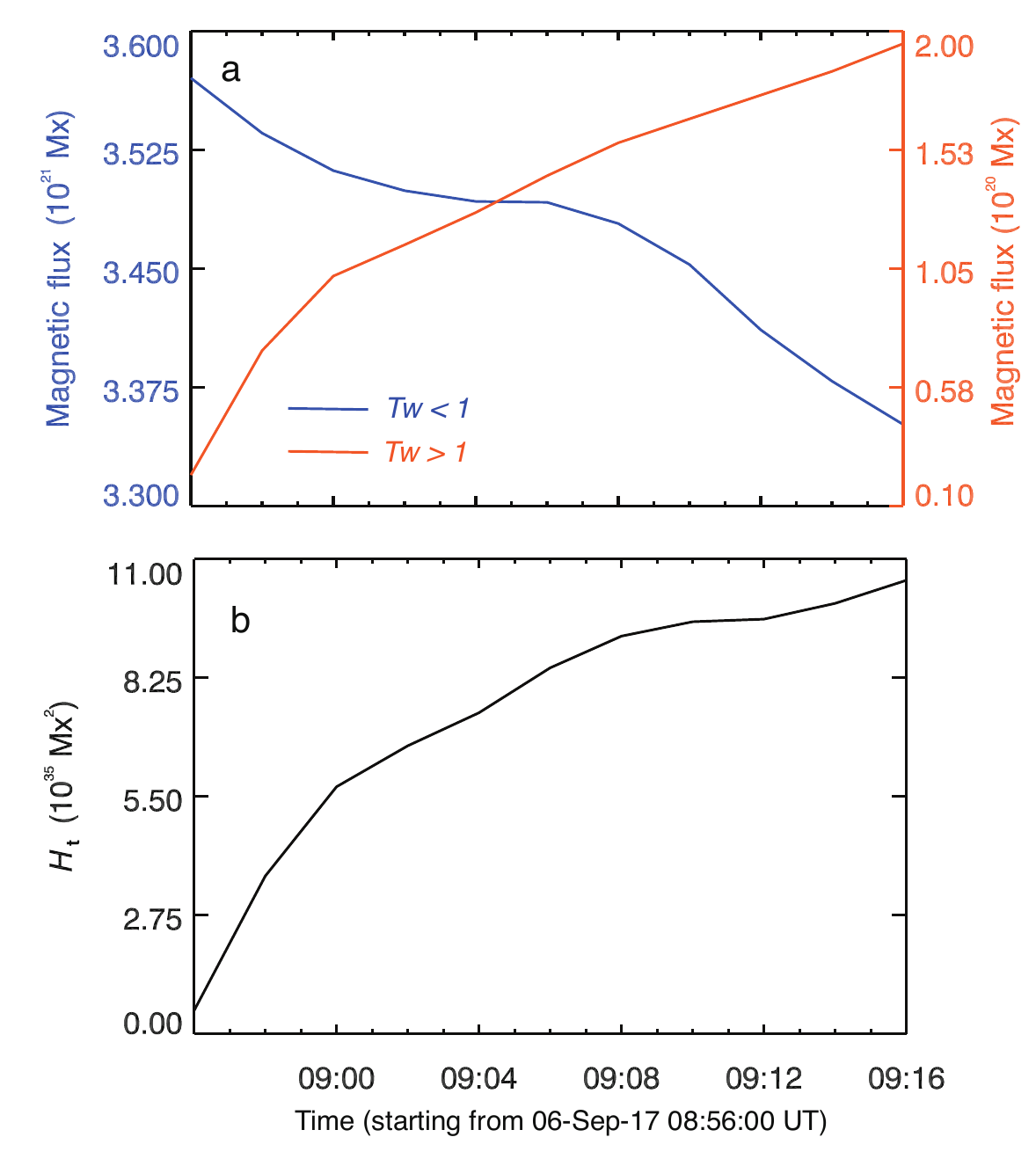}
\caption{Temporary evolution of (a) magnetic flux and (b) magnetic helicity of the twisted field lines. The blue and red lines in panel (a) correspond to the fluxes associated with $T_{w}<1$ and $T_{w}>1$, respectively. The magnetic helicity of the flux rope in panel (b) is estimated from the regions of $T_{w}>1$.}
\label{figure10}
\end{figure}

To quantify the accumulation of the twisted field lines during the confined eruption, following \citet{Liu2019} and \citet{Price2019}, we calculate the magnetic fluxes ($\int \boldsymbol{B_{T}} \cdot \boldsymbol{ds}$) with $|T_{w}|<1$ and $|T_{w}|>1$ (Figure~\ref{figure10}a), and the magnetic helicity of twisted field lines with $|T_{w}|>1$ using $H_{t}=|T_{w}|\phi^{2}$ (Figure~\ref{figure10}b), based on the $T_{w}$ distribution in the bottom plane. It is found that some weakly twisted field lines ($T_{w}<1$) are turned into more twisted ones ($T_{w}>1$) during the confined eruption, corresponding to an increase in the magnetic helicity of twisted field lines. This implies that there is an accumulation in twisted field lines during the confined eruption, which may feed the follow-up eruptions. Recently, \citet{Hassanin2022} proposed a model to explain the sequence of a confined and subsequent successful eruptions. They demonstrated that flux cancellation, acting on the post-reconnection loops of the preceding confined eruption, can form a new flux rope. Our simulation results suggest that the accumulation of twisted fluxes can occur during the confined eruption as well. In summary, our simulation provides a 3D dynamic evolution for the formation of twisted field lines during a confined eruption, which may serve as the embryo for subsequent successful eruption. 

\subsection{Comparison with previous numerical modellings}\label{cp}
As introduced in Section~\ref{sec:obs}, some numerical models have been extensively employed to investigate the evolution of this active region and associated solar flares \citep{Price2019, Moraitis2019, Liu2019, Inoue2021}. Hence, to evaluate the effectiveness of our newly developed data-driven model in reproducing the observations, it is essential to conduct a comparison between our simulation results and previous works.

First, we conducted a comparison of the magnetic helicity and energy budgets, as illustrated in Figure~\ref{figure3}. The trends and magnitudes in our simulation almost follow the results presented in the TMF simulation carried out by \citet{Price2019} for this active region. For instance, by 2017 September 5, our and their results exhibit a similarity in the helicity ratio, approximately at 0.15. However, the free energy ratio in our simulation is slightly higher than that of \citet{Price2019}. This discrepancy could be attributed to the difference in the initial condition (such as the starting time) and the employed data-driven boundaries ($v$-$B$ driven or $E$-driven). Additionally, the metrics in our simulation are also found to be in agreement with the results obtained from the continuous NLFFF extrapolations showcased in \citet{Moraitis2019}.

In addition to the long-term evolution of the active region, the eruption process also exhibits some similarities. For example, we identified two null-point reconnection sites at the eruption onset, as shown in Figure~\ref{figure7}. These topological structures have also been found in previous model results, and demonstrated to play a crucial role in initiating the eruption \citep{Mitra2018, Price2019, Bamba2020, Zou2020, Inoue2021, Yamasaki2021, Daei2023}. Further insights into the impacts of these topological structures on the onset process of this flare can be found in the discussion by \citet{Inoue2021}. Furthermore, the footpoints and the shape of MFR1 closely align with those of the confined flux ropes \citep{Inoue2021, Daei2023}. In addition, the magnetic structure of the newly-formed MFR3 at the end of our simulation (09:16~UT) resembles that of the flux rope constructed by the NLFFF extrapolation at 09:22~UT \citep{Liu2018}. The similarity in magnetic topological structures with prior results validates the soundness of our data-driven model in reproducing the solar eruption.

To conclude, the consistency between our simulation and prior observational data-based modelings \citep{Price2019, Inoue2021, Daei2023} for this active region demonstrates the rigidity of our newly developed data-driven model. Nevertheless, our data-driven modeling also presents some unique advances, wherein the combination of the TMF and MHD modelings enhance the ability to capture both the long-term buildup process and the following drastic eruptions. Moreover, the adoption of the thermodynamic MHD model enables us to investigate the evolution of thermal properties during the eruption.

It should be noted that the magnetic systems experience a non-smooth transition when we switch from the TMF model to MHD model, as evidenced by the rising motion of the flux rope MFR1 at the very beginning of the MHD model (see Figure~\ref{figure6}). Several factors are responsible for it. First, the final magnetic field in the TMF model is not force-free, and the nontrivial residual Lorentz force drives the flux rope to rise, as demonstrated by \citet{Afanasyev2023}. Indeed, the force-free metric ($\sigma_{J}$, \citeauthor{Wheatland2000}~\citeyear{Wheatland2000}) of our final TMF magnetic fields is $\sim$0.35, which is higher than that in our NLFFF model \citep{Guo2016b, Guo2019, Zhong2023}. Second, the switch from the TMF to MHD model inevitably introduces numerical errors, despite the employment of similar numerical schemes. In addition, the data-driven boundary is employed in the MHD simulation, where the colliding shearing facilitates the flux rope to rise, as demonstrated in \citet{Torok2018}.

\section{Summary}\label{sec:summary}

In this paper, we performed a data-driven simulation to investigate the evolution of NOAA Active Region 12673 and its associated confined X2.2 flare on 2017 September 6. The initial magnetic field condition is derived from a potential field model, and the subsequent evolution of the coronal magnetic field is fully driven by the photospheric magnetograms and velocity field derived from observations. We employed a hybrid data-driven model to simulate the development of this active region, including a TMF model to simulate the buildup of nonpotential energy and a thermodynamic MHD model to explore the subsequent eruption. Our simulations successfully captured the formation and eruption process of a flux rope, reproduced many observational features, and provided insights into underlying physical mechanisms. The key findings are summarized as follows:

\begin{enumerate}

\item{Our simulation results exhibited comparability with observations in several aspects. First, the simulated magnetic flux rope resembles the observed hot channel very well (Figure~\ref{figure4}b). Second, the observed S-shaped structure during the X2.2 flare is reproduced by our numerical model through the simulated QSLs and the synthesized EUV image (Figure~\ref{figure8}). Given all of these, we believe that our data-driven model possesses the capacity to reproduce the principal observational features and the underlying physical processes, to a certain extent.}

\item{Our data-driven model simulates the formation process of the flux rope before the eruption, along with a substantial increase in the magnetic free energy and current-carrying magnetic helicity (Figure~\ref{figure3}). Moreover, our simulation results suggest that collisional shearing and its consequent flux cancellation play a pivotal role in forming a twisted flux rope (Figure~\ref{figure4}).}

\item{Our data-driven model successfully reproduced the dynamic evolution of a confined eruption, enabling us to investigate the underlying physical mechanisms dominating failed eruptions. We found that the deformation of the flux rope can lead to a variation of the orientation in the external magnetic fields, transferring from the poloidal to toroidal direction, thereby leading to an increase in the downward toroidal-field tension force. As such, we suggest that the toroidal-field tension force may be also responsible for the failed eruption, particularly for events where the flux rope rotates significantly.}

\item{Our simulation results indicate that there may be a buildup of twisted fluxes during some confined flares. We found that some weakly twisted fluxes ($T_{w}<1$) are turned into more twisted fluxes ($T_{w}>1$) during the confined eruption, along with an increase in the helicity of the twisted flux rope. This implies that twisted fluxes could accumulate during confined eruptions, potentially breeding the subsequent strong eruptive flares.}

\end{enumerate}

\acknowledgments
We thank the anonymous referee for detailed and insightful suggestions which helped to improve the quality of this paper. J.H.G.\ acknowledges C. Wang for valuable discussions. The SDO data are available courtesy of NASA/SDO and the AIA and HMI science teams. This research was suppoted by the National Key R\&D Program of China (2020YFC2201200 and 2022YFF0503004), NSFC (12127901, 11961131002, 11773016, and 11533005). J.H.G. was supported by the China Scholarship Council under file No.\ 202206190140. S.P.\ acknowledges support from the projects C14/19/089  (C1 project Internal Funds KU Leuven), G.0B58.23N  (FWO-Vlaanderen), SIDC Data Exploitation (ESA Prodex-12), and Belspo project B2/191/P1/SWiM. Y.H.Z.\ acknowledges funding from Research Foundation - Flanders FWO under the project number 1256423N. The numerical calculations in this paper were performed in the cluster system of the High Performance Computing Center (HPCC) of Nanjing University.

\bibliography{ms}{}
\bibliographystyle{aasjournal}

\end{document}